\documentclass[preprint2]{aastex}

\usepackage{natbib}
\usepackage{lscape}
\usepackage{rotating}
\usepackage{amsmath}
\textheight=\baselinestretch\textheight
\ifdim\textheight>25.2cm\textheight=25.0cm\fi

\topskip\baselineskip
\maxdepth\baselineskip
\let\tighten=\relax
\let\tightenlines=\tighten

\newcommand{\cmmthree}{\mbox{cm$^{-3}$}}

\newcommand{\degree}{\mbox{$^{\circ}$}}

\newcommand{\uchii}{\mbox{UCH\,{\scriptsize II}}}
\newcommand{\hii}{\mbox{H\,{\scriptsize II}}}
\newcommand{\crn}{\mbox{\small CORNISH}}
\newcommand{\gli}{\mbox{\small GLIMPSE}}
\newcommand{\mip}{\mbox{\small MIPSGAL}}

\slugcomment{To appear in ApJ Supplement.}

\shorttitle{[The $\crn$ Source Catalogue}
\shortauthors{[C.~R.~Purcell et al.}

\begin{document}

\title{The Co-ordinated Radio and Infrared Survey for High-Mass Star
  Formation - II. Source Catalogue} 

\author{
C.~R.\,Purcell\altaffilmark{1,2,3,}$^{\dagger}$,
M.~G.~Hoare\altaffilmark{1}, 
W.~D.~Cotton\altaffilmark{4}, 
S.~L.~Lumsden\altaffilmark{1}, ,
J.~S.~Urquhart\altaffilmark{1,5,6}
C.~Chandler\altaffilmark{7},
E.~B.~Churchwell\altaffilmark{8},
P.~Diamond\altaffilmark{2,5},  
S.~M.~Dougherty\altaffilmark{9}, 
R.~P.~Fender\altaffilmark{10}, 
G.~Fuller\altaffilmark{2},
S.~T.~Garrington\altaffilmark{2},
T.~M.~Gledhill\altaffilmark{11}, 
P.~F.~Goldsmith\altaffilmark{12}, 
L. Hindson\altaffilmark{5,11,}, 
J.~M.~Jackson\altaffilmark{13},
S.~E.~Kurtz\altaffilmark{14}, 
J.~Mart\'i\altaffilmark{15},
T.~J.~T.~Moore\altaffilmark{16}, 
L.~G.~Mundy\altaffilmark{17},
T.~W.~B.~Muxlow\altaffilmark{2},
R.~D.~Oudmaijer\altaffilmark{1},
J.~D.~Pandian\altaffilmark{18}, 
J.~M.~Paredes\altaffilmark{19}, 
D.~S.~Shepherd\altaffilmark{7,20},
S.~Smethurst\altaffilmark{2},
R.~E.~Spencer\altaffilmark{2}, 
M.~A.~Thompson\altaffilmark{11},
G.~Umana\altaffilmark{21} and
A.~A.~Zijlstra\altaffilmark{2}}

\altaffiltext{$\dagger$}{\bf C.R.Purcell@leeds.ac.uk}
\altaffiltext{1}{School of Physics \& Astronomy, E.C. Stoner Building,
  University of Leeds, Leeds LS2 9JT, UK}
\altaffiltext{2}{Jodrell Bank Centre for Astrophysics, The Alan Turing
  Building, School of Physics and Astronomy, The University of
  Manchester, Oxford Rd, Manchester, M13 9PL, UK}
\altaffiltext{3}{Sydney Institute for Astronomy (SiFA), School of
Physics, The University of Sydney, NSW 2006, Australia}
\altaffiltext{4}{National Radio Astronomy Observatory, 520 Edgemont Road,
  Charlottesville, VA 22903-2475, USA}
\altaffiltext{5}{CSIRO Astronomy and Space Science, PO BOX 76, Epping,
  NSW 1710, Australia}
\altaffiltext{6}{Max-Planck-Institut fu\"{r} Radioastronomie, Auf dem
  Hu\"{g}el 69, D-53121 Bonn, Germany}
\altaffiltext{7}{National Radio Astronomy Observatory, Array Operations
  Center, P.O. Box O, 1003 Lopezville Road, Socorro, NM 87801-0387, USA}
\altaffiltext{8}{The University of Wisconsin, Department of Astronomy,
  475 North Charter Street Madison, WI 53706, USA}
\altaffiltext{9}{National Research Council of Canada, Herzberg
  Institute for Astrophysics, Dominion Radio Astrophysical 
  Observatory, PO Box 248, Penticton, British Columbia V2A 6J9, Canada}
\altaffiltext{10}{School of Physics and Astronomy, University of
  Southampton, Southampton SO17 1BJ}
\altaffiltext{11}{Science and Technology Research Institute, University of
     Hertfordshire, College Lane, Hatfield AL10 9AB, UK}
\altaffiltext{12}{Jet Propulsion Laboratory, 4800 Oak Grove Drive,
  Pasadena, California 91109, USA}
\altaffiltext{13}{Astronomy Department, Boston University, 725
  Commonwealth Avenue, Boston, MA 02215, USA}
\altaffiltext{14}{Centro de Radioastronom\'ia y Astrof\'isica, UNAM -
  Morelia, Apartado Postal 3-72, C.P. 58090 Morelia, Michoacan, Mexico}
\altaffiltext{15}{Departamento de F\'isica, EPSJ, Universidad de
  Ja\'en, Campus Las Lagunillas s/n, Edif. A3, 23071 Ja\'en, Spain}
\altaffiltext{16}{Astrophysics Research Institute, Liverpool John
  Moores University, Twelve Quays House, Egerton Wharf, Birkenhead
  CH41 1LD, UK}
\altaffiltext{17}{Department of Astronomy, University of Maryland
  College Park, MD 20742-2421, USA}
\altaffiltext{18}{Institute for Astronomy, 2680 Woodlawn Drive,
  Honolulu, Hawaii 96822-1839, USA}
\altaffiltext{19}{Departament d'Astronomia i Meteorologia and Institut
  de Ciencies del Cosmos (ICC), Universitat de Barcelona (UB/IEEC),
  Mart\'i Franqu\`s 1, 08028 Barcelona, Spain}
\altaffiltext{20}{Square Kilometer Array - South Africa, 3rd floor,
  The Park, Park Rd, Pinelands, Cape Town, 7405 Western Cape, South Africa}
\altaffiltext{21}{INAF – Osservatorio Astrofisico di Catania, via
  S. Sofia 78, 95123 Catania, Italy}

\begin{abstract}
The $\crn$ project is the highest resolution radio continuum survey of
the Galactic plane to date. It is the 5\,GHz radio continuum part of a
series of multi-wavelength surveys that focus on the northern $\gli$ region
($10^{\circ}<l<65^{\circ}$), observed by the {\it Spitzer}
satellite in the mid-infrared. Observations with the Karl G. Jansky
Very Large Array (VLA) in B and BnA configurations have yielded a
1.5$''$ resolution Stokes~{\it I} map with a root-mean-squared noise
level better than $0.4$\,mJy\,beam$^{-1}$. Here we describe the
data-processing methods and data characteristics, and present a new,
uniform catalogue of compact radio-emission. This includes an implementation of 
automatic deconvolution that provides much more reliable imaging than
standard {\scriptsize CLEAN}ing. A rigorous investigation of the noise
characteristics and reliability of source detection has been carried
out. We show that the survey is optimised to detect emission on size
scales up to $14''$ and for unresolved sources the catalogue is more
than 90 percent complete at a flux density of 3.9\,mJy. We have
detected 3,062 sources above a $7\sigma$ detection limit and present
their ensemble properties. The catalogue is highly reliable away from
regions containing poorly-sampled extended 
emission, which comprise less than two percent of the survey
area. Imaging problems have been mitigated by down-weighting the
shortest spacings and potential artefacts flagged 
via a rigorous manual inspection with reference to the {\it Spitzer}
infrared data. We present images of the most common source
types found: $\hii$ regions, planetary nebulae and radio-galaxies. The
$\crn$ data and catalogue are available online at
http://cornish.leeds.ac.uk.
\end{abstract}

\keywords{catalogues -- ISM: H\,{\scriptsize II} -- radio continuum: general 
  -- radio continuum: ISM  -- surveys -- techniques: image processing}


\section{Introduction}
The observed progression of massive star formation, from cold
collapsing core to young OB clusters, is largely understood via
observations of discrete examples that have been ordered into an 
evolutionary sequence. Key to separating objects of different age and
type are measurements of their spectral energy distributions (SEDs) at
sub-millimetre, infrared and radio wavelengths. 

The {\it Spitzer} $\gli$ (Galactic Legacy Infrared Mid-Plane Survey
Extraordinaire) programme is the first of a number of sensitive infrared
surveys covering the inner Galactic plane at high resolution and in an
unbiased manner \citep{Churchwell2009}. The northern half of $\gli$
covers the region $10^{\circ}<\,l\,<65^{\circ},~|b|<1^{\circ}$ at
wavelengths spanning 3.6\,$\micron$\,-\,8.0\,$\micron$, which preferentially
selects warm and dusty 
embedded sources. The companion {\it Spitzer} $\mip$ survey
\citep{Carey2009} has imaged the same region at 24\,$\micron$ and 70\,$\micron$
(where the bulk of the energy from massive young stellar objects is
emitted) and is hence sensitive to cooler and more deeply embedded
young stellar objects. Most recently, the {\it Herschel} Infrared
Galactic Plane survey (Hi-GAL, \citealt{Molinari2010a}) is
delivering the most comprehensive survey of embedded objects to date. With
observations in six far-infrared bands between 70\,$\micron$ and 500\,$\micron$,
Hi-GAL samples the peak of the star-forming SED and covers the
northern $\gli$ region out to $l=60$. Completing the infrared picture
of Galactic star formation is the
UKIDSS\footnote{http://www.ukidss.org} project (UK IR Deep Sky Survey,
\citealt{Lawrence2007}). A subset of UKIDSS (the Galactic Plane Survey,
\citealt{Lucas2008}) has observed the northern $\gli$ region in the
near-infrared J, H and K bands and is sensitive to objects down to 18th
magnitude. The combined data from these surveys
are driving the detailed characterisation of the Galactic population
via their infrared colours (e.g., \citealt{Robitaille2007},
\citealt{Arvidsson2010}, \citealt{Smith2010},
\citealt{Wright2010},\citealt{Mottram2011}). A complementary picture of the  
molecular and atomic interstellar medium is being provided by the
BU-FCRAO Galactic Ring Survey for CO \citep{Jackson2006} and the VLA
Galactic Plane Survey (VGPS) for H\,{\scriptsize I}
\citep{Stil2006}. Similarly, the ongoing Isac Newton Telescope
Photometric Survey of the Northern Galactic Plane (IPHAS) 
\citep{Drew2005} probes H\,$\alpha$ in emission towards nebulae, and in
both absorption and emission towards stars. The UKIRT Wide Field
Infrared Survey for H$_2$ (UWISH2, \citealt{Froebrich2011}) also
covers the same $\gli$ region in molecular hydrogen (2.122\,$\micron$
line) highlighting regions of shocked or fluorescently excited
molecular gas (T$\approx2000$\,K, $n_{\rm H_2}>10^3\,\cmmthree$).

Conspicuous by its absence is a comparable radio continuum survey
for compact ionised gas. Previous surveys are either targeted at
individual sources selected via infrared colours (e.g.,
\citealt{Wood1989}, \citealt{Kurtz1994}, \citealt{Urquhart2009}) or are
limited in their resolution and sky-coverage (e.g.,
\citealt{Becker1994}, \citealt{White2005}). From a
star formation perspective, the presence or absence of free-free
emission is vital to distinguish the more evolved ultra-compact
$\hii$ ($\uchii$) regions from their younger
counterparts with similar thermal SEDs \citep{Urquhart2009,Urquhart2011}. 
The sheer number density of sources in the near and mid-infrared
surveys necessitates complementary data at similarly high resolution
to enable the full science potential to be fulfilled. This is
particularly true in highly clustered star forming regions.
It is important that any radio-continuum survey for $\uchii$ regions
be carried out at relatively high frequencies ($\ge$5\,GHz) 
where thermal free-free emission is optically thin with a spectral
index of $S_{\nu}\propto\nu^{-0.1}$. At lower frequencies the
spectrum becomes optically thick with $S_{\nu}\propto\nu^{2}$. 
High-frequency observations hence confer a signal-to-noise advantage
and probe the structure of the ionised gas at all depths in
$\uchii$ regions. We note that even at $\nu=5$\,GHz we will
be insensitive to a population of young and compact $\hii$ regions: the
so-called Hyper-compact $\hii$ (HC$\hii$) regions 
(see \citealt{Sewilo2011} and references therein). These objects have
greater emission measures than $\uchii$ regions and the
turnover frequency from optically thick to thin occurs at high
frequencies. 

No previous radio survey of the Galactic plane has similar resolution
and coverage to the {\it Spitzer} $\gli$ survey. A number of 
single dish surveys have been conducted at 5\,GHz (e.g.,
\citealt{Altenhoff1979}), however, their arcminute resolution is quite
low, compared to the arcsecond resolution of {\it Spitzer}. Most
interferometric surveys have been carried out at a frequency of
1.4\,GHz (e.g., the NRAO VLA Sky Survey, \citealt{Condon1998}) except
for the catalogues of \cite{Becker1994}, 
\cite{Giveon2005} and \cite{White2005}, who surveyed the inner
Galactic plane ($-10^{\circ}\,<\,l\,<\,42^{\circ},~|b|\,<\,0.4^{\circ}$) at
5\,GHz. These three surveys are published as the Multi-Array Galactic
Plane Imaging Survey
(MAGPIS)\footnote{http://third.ucllnl.org/gps/}. They used the Very
Large Array (VLA) in C and D-configurations, which deliver a
relatively large beam (4$''$$\times$9$''$) and the total survey area
only covers 26 percent of the northern $\gli$ region.  

The $\crn$ (Co-Ordinated Radio `N' Infrared Survey for High-mass star
formation) project delivers a uniform, sensitive and
high-resolution radio survey of the northern $\gli$ region to address
key questions in high-mass star formation, as well as many other areas
of astrophysics. In addition to $\uchii$ regions, the
$\crn$ survey detects many other radio-bright objects, including planetary
nebulae, ionised winds from evolved massive stars, non-thermal
emission from active stars, active Galactic nuclei and radio
galaxies. The full rationale behind the survey design and the
scientific motivation is presented in an accompanying paper, \citep{Hoare2012}.


\begin{figure*}
  \centering
  \includegraphics[angle=90, width=17.7cm, trim=0 0 0 0]{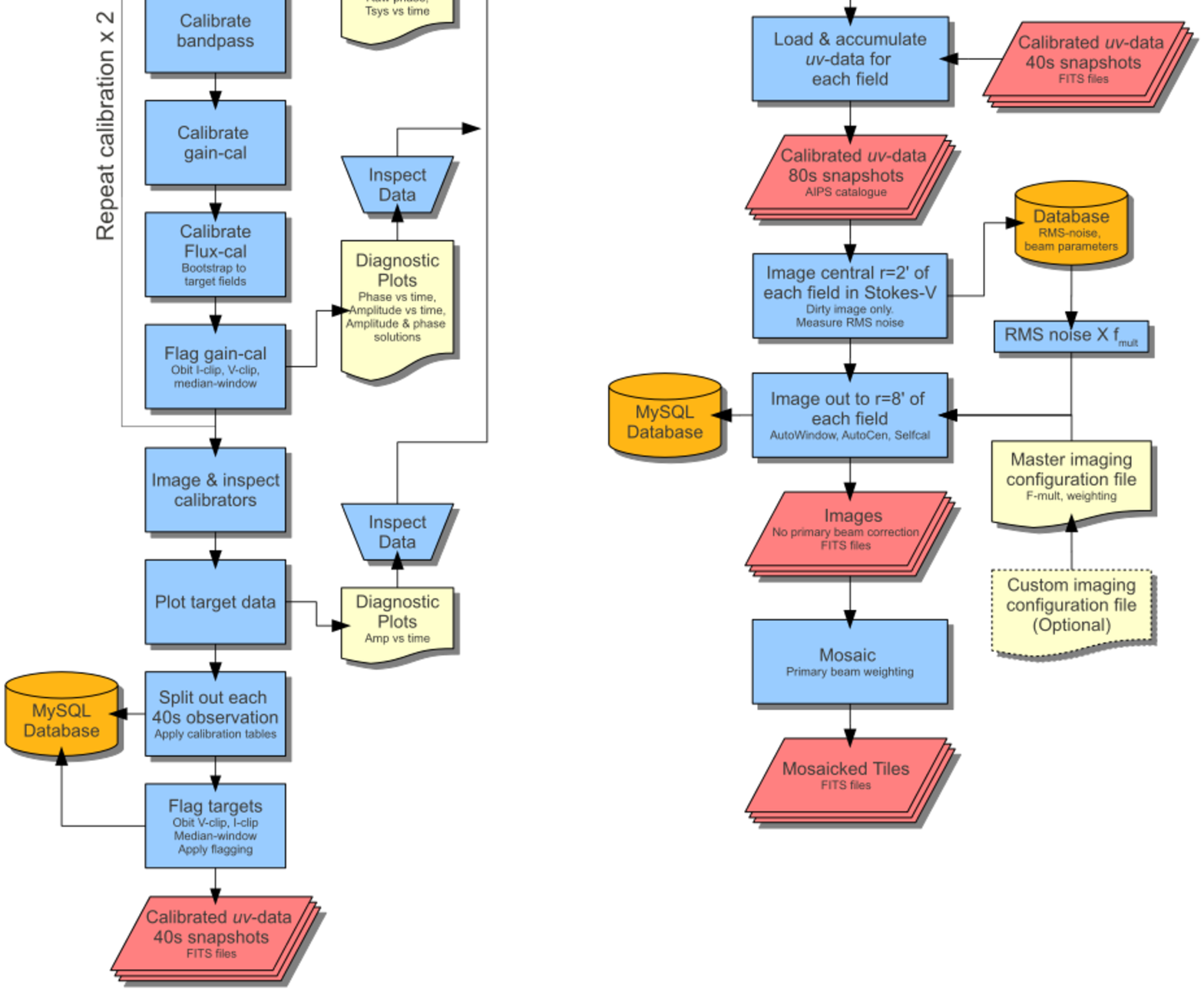}
  \caption{\small A graphical illustration of the $\crn$ calibration and imaging
    pipeline. See the text of Section~\ref{sec:pipeline} for more detail.}
  \label{fig:pipeline_flow}
\end{figure*}
\begin{table*}    
\caption{Details of the $\crn$ observational epochs.\label{tab:obs}} 
\begin{small}
\begin{tabular}{l@{~}r@{\,}c@{\,}l@{~~~}r@{\,}c@{\,}lr@{\,}c@{\,}l@{~~}l@{~~~}l}
\tableline
\tableline
Epoch & \multicolumn{3}{c}{Dates}                                  & \multicolumn{3}{c}{Dec. Range}                  & \multicolumn{3}{c}{{\it l} Range}          & Config. & Notes         \\
\tableline                                                                                                                                                                                                        
I    & 2006 Jul\, 12$^{\rm th}$ &$\rightarrow$ & 2006 Sep 16$^{\rm th}$    & $-$10.5$^{\circ}$&$\rightarrow$&$+$14.2$^{\circ}$ & 21.1$^{\circ}$&$\rightarrow$&48.9$^{\circ}$  & B      & VLA antennas only, storms.    \\
II   & 2007 Sep 28$^{\rm th}$ &$\rightarrow$ & 2007 Oct ~\,6$^{\rm th}$  & $-$20.8$^{\circ}$&$\rightarrow$&$-$14.9$^{\circ}$ & 10.0$^{\circ}$&$\rightarrow$&16.1$^{\circ}$  & BnA    & VLA + EVLA antennas, low Dec. \\
IIIa & 2007 Oct 27$^{\rm th}$ &$\rightarrow$ & 2008 Feb ~\,4$^{\rm th}$  & $-$14.9$^{\circ}$&$\rightarrow$&$-$10.5$^{\circ}$ & 16.1$^{\circ}$&$\rightarrow$&21.1$^{\circ}$  & B      & VLA + EVLA antennas.          \\
IIIb & 2007 Oct 27$^{\rm th}$ &$\rightarrow$ & 2008 Feb ~\,4$^{\rm th}$  & $+$14.2$^{\circ}$&$\rightarrow$&$+$29.1$^{\circ}$ & 48.9$^{\circ}$&$\rightarrow$&65.5$^{\circ}$  & B      & VLA + EVLA antennas.          \\
\tableline
\end{tabular}
\end{small}
\tablecomments{The properties of the data differ in the combination of
  antenna types included in the array, the configuration of the array,
  the weather experienced and the declination range observed. Unless
  otherwise noted the weather during the observations was reasonable.} 
\end{table*}

\section{Observations}\label{sec:observations} 
$\crn$ covers the 110 square degrees of the northern $\gli$ region
($10^{\circ}\,<\,l\,<\,65^{\circ},~|b|\,<\,1^{\circ}$) using the VLA
in B and BnA configurations at 5\,GHz. The combination of array
configuration and observing frequency results in a $\sim1.5''$ synthesised
beam within a $8.9'$ field of view, corresponding to the full-width
half-maximum (FWHM) primary beam. With a total integration time of 80
seconds per pointing, the root-mean-squared (RMS) noise in the images is
better than 0.4 mJy\,beam$^{-1}$ - sufficient to detect an unresolved
$\uchii$ region around a B0 star on the far edge of the
Galaxy (16\,kpc, \citealt{Kurtz1994}).

$\crn$ observations of the northern $\gli$ region were conducted using
the VLA during the 2006 and 2007/2008 observing
seasons. The observations fall naturally into the epochs presented
in Table~\ref{tab:obs}, which are distinguished by the combinations of
array configuration used, inclusion or exclusion of upgraded EVLA
antennas, declination ranges observed and weather conditions
experienced. We show later that data from each epoch have unique
properties.

To facilitate scheduling the target area was divided into 42 blocks each
corresponding to eight hours of observations per day. Block contain
between 180 and 220 fields arranged in rows of equal right-ascension
on a hexagonal pointing grid.  Individual fields were observed as two
45 second `snapshots' separated by $\sim4$ hours in time, maximising the
{\it uv}-coverage and minimising the elongation of the synthesised
beam. The telescope was advanced along each row ($\sim20$ fields) integrating
for 45 seconds on each pointing position, before observing a secondary
calibrator (one of 1832-105, 1856+061 or 1925+211) for two
minutes and then continuing to the next row. Including overheads the
secondary calibrators were observed with a cadence of twenty
minutes. Fields at declinations greater than $-15\degree$ were
observed using the VLA's B configuration while  fields at lower
declinations were observed using the BnA configuration, which is
designed to compensate for beam distortion at low elevations. 

To allow imaging of the widest possible field of view without
bandwidth-smearing the observations were carried out in
pseudo-spectral line mode. The two 25\,MHz wide spectral windows
(also known as intermediate frequencies, or IFs) of the VLA correlator
were tuned to adjoining frequency bands centred on 5\,GHz. Each window
was sampled by eight 3.1 MHz channels, degrading the peak response by
only a few percent at the edge of the $8.9'$ primary beam. Due to hardware
limitations only the RR and LL polarisations were recorded,
meaning that linear polarisation information is not available in the
$\crn$ data. 

During both $\crn$ observing seasons significant engineering works
were underway to upgrade the VLA to the next generation instrument: the
Expanded VLA (EVLA). In 2006 between 
two and six antennas were missing from the array as they were being
refurbished with new receivers and electronics to convert them to the
EVLA design. By the start of the second season of $\crn$ observations 
(September 2007), almost half the array was comprised of EVLA antennas
and the instrument was operating in a transition mode. Over the season
VLA antennas were progressively removed from the active array and
substituted by EVLA antennas. The EVLA 
antennas conferred the advantage of enhanced sensitivity, but at the
same time were untested and prone to software and hardware
problems. Special care was needed to properly calibrate VLA-EVLA
baselines and to ensure that the EVLA data were properly flagged. 
As part of the upgrade, the venerable Modcomp-based VLA control
systems was also replaced in mid 2007 with new software running under
Linux. Taken as a whole, the $\crn$ data required close inspection and
vigilance during post-processing.


\section{The data reduction pipeline}\label{sec:pipeline}
The raw $\crn$ dataset consists of 9,349 pointing positions, each
of which was observed twice. A manually guided data-reduction
procedure was considered too labour-intensive to use on such a large
volume of data, hence a semi-automatic pipeline was developed with the
control parameters tuned to the average observation. This approach has
the advantage of applying uniform  processing over the majority of the
survey area, while still allowing manual intervention in a minority of
special cases (e.g., fields with complicated emission structures, or
very bright sources).

The pipeline was implemented in the {\it python} language and made use
of the {\scriptsize ObitTalk} module to interface directly with the
NRAO\footnote{The National Radio Astronomy Observatory is a facility
  of the National Science Foundation operated under cooperative
  agreement by Associated Universities, Inc.}
{\scriptsize AIPS}\footnote{http://www.aips.nrao.edu/} and 
{\scriptsize
  OBIT}\footnote{http://www.cv.nrao.edu/$\sim$bcotton/Obit.html}
data-reduction 
packages. The $\crn$ pipeline utilised a MySQL database to record
meta-data and perform bookkeeping operations during the reduction
procedure. 

Figure \ref{fig:pipeline_flow} illustrates the pipeline logic,
which is broken up into calibration and imaging stages. In the
following sections we describe each of the stages in detail.


\subsection{Calibration and flagging}\label{sec:calibration}
Raw data from the telescope were corrected for atmospheric opacity
using phase monitor data and written to an {\scriptsize AIPS}-format
{\it uv}-file in spectral-line mode by the {\scriptsize AIPS} task
{\scriptsize FILLM}. Each eight-hour block of observations was first
inspected by eye  and {\it uv}-visibilities with large phase scatter, 
errant amplitudes or system-temperature spikes were flagged out of the
dataset. Gross errors in the data, such as bad antennas, IFs or
polarisations were also flagged out at this stage. It was necessary to
edit out the first five seconds of data from each pointing to allow for
antenna settling time, reducing the on-source integration time from
45-sec to 40-sec. All 
manual flagging parameters were written to a master flag list, which
was automatically 
applied upon restarting the pipeline. Care was taken here that the
primary flux calibrators contained only good data.

The shapes of the VLA and EVLA pass-bands are different enough that a
six percent closure error has been measured on EVLA-VLA baselines in
continuum modes using 50-MHz bandwidths\footnote{See
http://www.vla.nrao.edu/astro/guides/evlareturn/ for details.}. This error is
expected to be larger at narrower bandwidths. Because we are operating in
pseudo-spectral line mode the issue was mitigated by performing
bandpass calibration (phase and amplitude) immediately 
after the initial flagging, and before any further calibration. Solutions
for the atmospherically and electronically induced changes in phase
and amplitude were then calculated using the standard {\scriptsize
  AIPS} {\scriptsize CALIB} task, operating on one 
of the three secondary calibrators. The data were bootstrapped on to an
absolute flux scale by comparing observations of the quasars 1331+305
(3C286) or 0137+331 (3C48) to their C-band model in {\scriptsize
  AIPS}. A global calibration table was produced, which could then be
applied to the whole block. 

After a first pass at calibration the {\scriptsize OBIT}
flagging tasks {\scriptsize AutoFlag} and {\scriptsize MednFlag} were
applied to  
each IF, polarisation and channel of the secondary calibrator
observations. {\scriptsize AutoFlag} edits out bad visibilities based on an
absolute maximum allowed value in Stokes~{\it I} or {\it V}. Radio-frequency
interference (RFI), e.g., from commercial broadcasting, is often
highly polarised and all visibilities with
Stokes~{\it V} amplitude greater than 2\,Jy were flagged as
bad.{\scriptsize MednFlag} applies a rolling median filter to each IF
and spectral channel. Visibilities were edited out if they had
Stokes~{\it I} values greater than five standard deviations from the
median, calculated in a $\sim50$ second time-window. A second pass at
calibration was then performed before  applying a similar flagging
procedure to the calibrated science observations on a field-by-field
basis. Finally, all calibration and flagging tables were applied to
the data and the individual forty-second pointings were split into
{\it uv}-FITS format files.

Meta-data associated with each observation (e.g., the pointing
centre co-ordinates and number of flagged visibilities) were
automatically saved in the MySQL database. The imaging procedure
subsequently queried this database when building a final mosaiced image.


\begin{figure}
  \centering
  \includegraphics[width=8.0cm, trim=20 0 40 30]{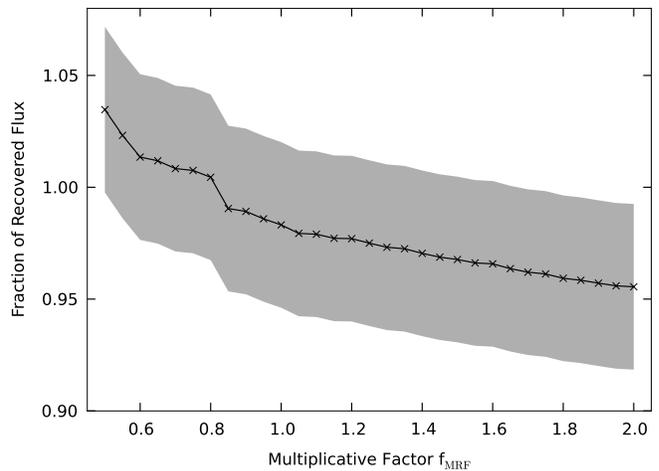}
  \caption{\small Fraction of recovered flux versus multiplicative factor
  $f_{\rm MRF}$ for an artificial source injected into the empty field
   18173$-$18247. The target `maximum residual flux' (MRF) driving the
   deconvolution procedure was set to RMS${\rm _V}\times f_{\rm MRF}$,
   where RMS${\rm 
  _V}$ is the root-mean-squared noise measured in Stokes~{\it V} dirty
  map. The amplitude of the RMS noise in the image expressed as a
  fraction of the recovered flux is illustrated by the grey-shaded
  area. For factors below $\sim0.8$ the field is over-cleaned, leading
  to significant artefacts in the image. Values less than 2.0 recover
  greater than 96 percent of the flux.} 
  \label{fig:clnflux_vs_mult}
\end{figure}

\subsection{Imaging}\label{sec:imaging}
Fields were imaged using the {\scriptsize OBIT} {\scriptsize Imager}
task, which performs imaging and deconvolution in a similar manner to
the {\scriptsize AIPS} {\scriptsize IMAGR} routine. {\scriptsize Imager}
automatically switches between the standard Cotton-Schwabb and SDI
deconvolution algorithms (\citealt{Schwab1984}, \citealt{Steer1984}),
referred to simply as `{\scriptsize CLEAN}' in the following 
discussion. In addition the task can be instructed to perform both
phase and amplitude self-calibration. {\scriptsize Imager} is a complex task
with many important input parameters which need careful tuning to result in a
scientifically useful image. Key amongst these are the maximum
residual flux, the threshold at which to begin self-calibration,
whether to perform both phase and amplitude self-calibration, and the
weighting function used. The `average' $\crn$ field was imaged using
a Briggs-robustness parameter of zero, which is a compromise between
natural (high-sensitivity) and uniform (high-resolution)
weighting. A minority of complex fields were treated as a special case
and imaged with a custom weighting scheme - see
Section~\ref{sec:img_extended}, below. Self-calibration was
performed on sources with peak fluxes greater than 30\,mJy\,beam$^{-1}$. During the
deconvolution process the restoring Gaussian beam was forced to be
circular and have a full-width half-maximum of $1.5''$. The cell-size
was set at $0.3''$, over-sampling the synthesised beam. We justify
these choices in the following sections.

\begin{figure*}
  \centering
  \includegraphics[width=18.0cm, angle=0, trim=0 0 0 0]{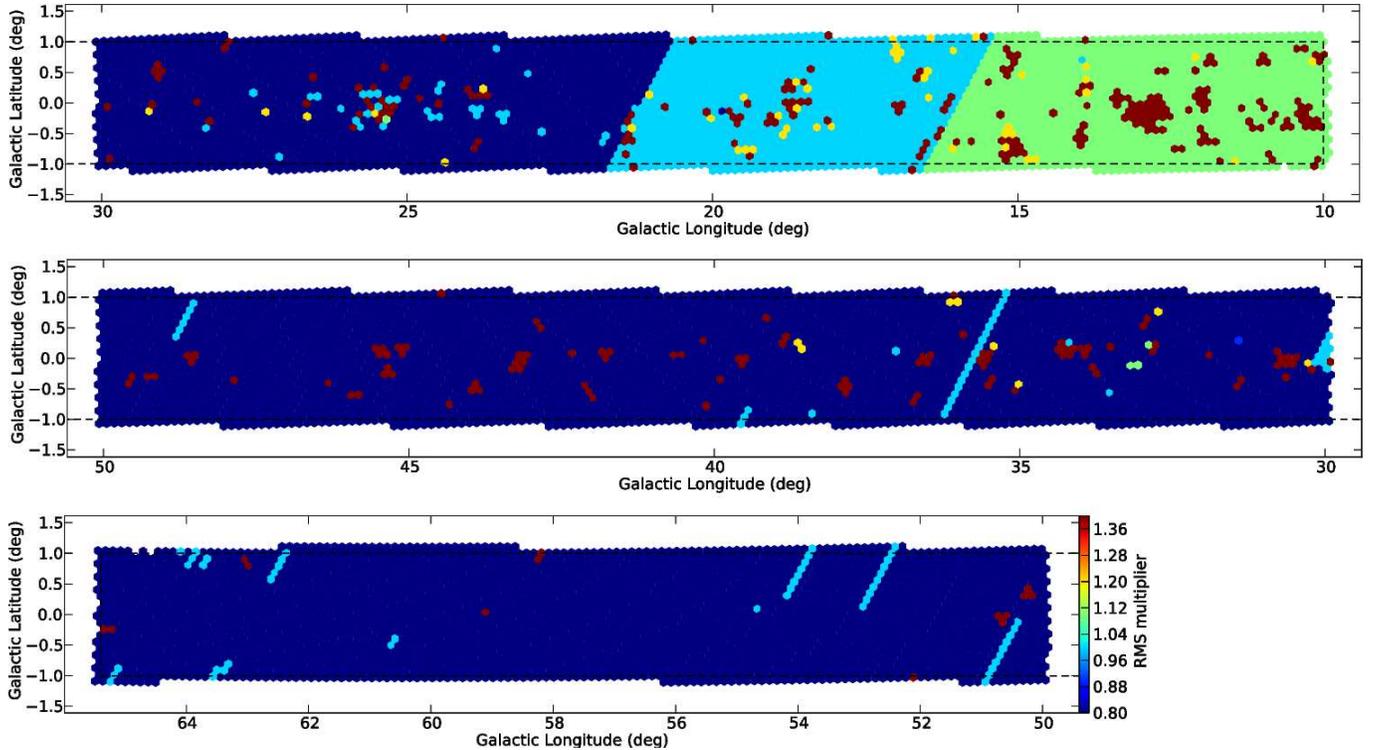}
  \caption{\small Map of the multiplier $f_{\rm MRF}$ used to control the
    deconvolution process on each field. Fields at lower declinations
    required higher values of $f_{\rm MRF}$ to compensate for the decrease in
    sensitivity which accompanied the lower observing
    elevations. Individual fields containing bright or complex
    structured emission also required higher $f_{\rm MRF}$
    values. Partial rows of fields (in R.A.) with different
    multipliers compared to their neighbours are due to repeat
    observations from later epochs. The row at
    $\delta=-16\degree55'03.13''$ ($13.3\degree<l<14.5\degree$) was
    observed only once and the $f_{\rm MRF}$ was increased to avoid
    over-cleaning.}
  \label{fig:fmrf}
\end{figure*}

{\scriptsize Imager} implements two algorithms which improve the
dynamic range and quality of the final image. Firstly, the
{\scriptsize
  AutoWindow}\footnote{http://www.aoc.nrao.edu/evla/geninfo/memoseries/evlamemo116.pdf} function dynamically places small ($<20$\,pixels in radius) clean
windows over regions of emission in the intermediate dirty map. The
validity of each window is assessed periodically during the clean
cycle and additional windows are created as necessary. The net effect
is to {\scriptsize CLEAN} only real emission and avoid {\scriptsize
  CLEAN}ing noise, resulting in a smaller clean bias. The {\scriptsize
  AutoCen}\footnote{http://www.aoc.nrao.edu/evla/geninfo/memoseries/evlamemo114.pdf}
algorithm regrids {\it uv}-data containing bright point sources, so
their peaks fall on a pixel centre. A point source at the centre of a pixel
can be represented as a single delta function, i.e., a single
clean-component, leading to a significant improvement in dynamic
range. In contrast, a bright point source offset from a 
pixel centre requires multiple clean components, both positive and negative,
to model its emission. This more complex model will inevitably suffer
from rounding errors on finite-precision computers, leading to flux
being scattered into the surrounding sky. The EVLA memos numbers
116 and 114 by Bill Cotton contain detailed descriptions of the
{\scriptsize AutoWindow} and {\scriptsize AutoCen} algorithms.


\subsubsection{Controlling the deconvolution algorithm}
One of the most critical control parameters for the imaging task is
the target maximum residual flux (MRF). The ideal value varies from
field to field depending on the weather conditions, individual antenna
system temperatures and the structure and strength of emission in the
field of view. To estimate the intrinsic sensitivity attainable we
imaged each field in Stokes~{\it V} and measured the RMS
noise. Sources with significant circularly polarised emission at
5\,GHz are rare (\citealt{Roberts1975}, \citealt{Homan2006}), so the
RMS noise measured from an {\it uncleaned} Stokes~{\it V} image is
expected to be comparable to the final {\scriptsize CLEAN}ed noise
level. Because the two polarisation beams of the VLA are not
co-aligned on the sky 
they give rise to a strong instrumental polarisation away from the
pointing centre. To compensate for this `beam
squint'\,\footnote{http://www.aoc.nrao.edu/evla/geninfo/memoseries/evlamemo113.pdf}
only the central $2'$ diameter portion of each field was imaged and
measured. Despite this we expect noise in the Stokes~{\it
  V} images to be higher than the ideal in their Stokes~{\it
  I} counterparts. In addition, the $1.5''$ restoring beam
applied to the {\it I} images is larger, on average, than the
unconstrained synthesised beam of the {\it V} images. This effectively 
smooths the noise in the Stokes~{\it I} images compared to the {\it V}.

\begin{figure}
  \centering
  \includegraphics[width=8.0cm, trim=20 20 50 40]{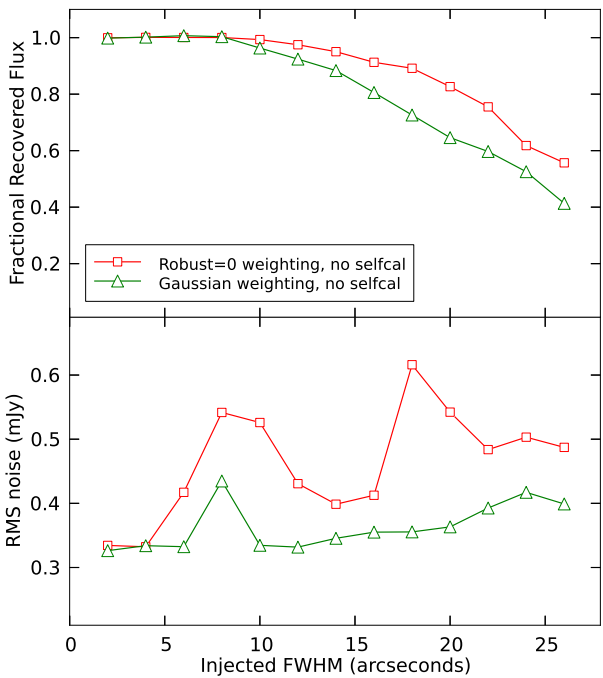}
  \caption{\small {\it Top panel:} Fractional recovered flux as a
    function of full-width half-maximum (FWHM) for an artificial 
    1\,Jy\,beam$^{-1}$ {\it peak} Gaussian source injected into an
    empty field. Data imaged with a Gaussian smoothed weighting is
    missing greater fractional flux compared to data imaged with
    robust\,=\,0 weighting. {\it Bottom panel:} Root-mean-square
    noise as a function of FWHM for 1\,Jy {\it flux density} Gaussian
    sources. The images made using the Gaussian smoothed weights tend
    to have lower and more stable noise properties.}
  \label{fig:missing_flux}
\end{figure}
\begin{figure}
  \centering
  \includegraphics[width=8.0cm, trim=20 20 50 40]{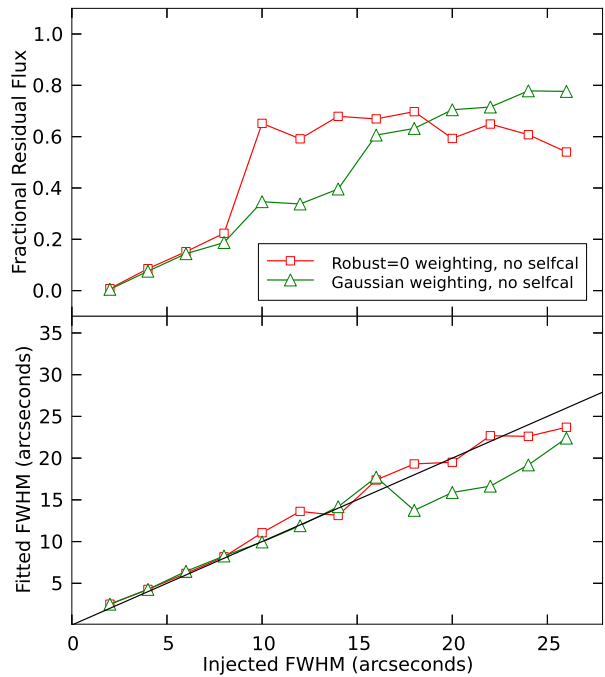}
  \caption{\small {\it Top panel:} Fractional absolute residual flux
    remaining after subtracting a model Gaussian source from the
    test image. Low values mean that the image is similar to the
    model while high values mean that there are significant differences
    in morphology. {\it Bottom panel:} Fitted versus injected
    FWHM. Sources with FWHM~$>$14$''$ are poorly imaged by the VLA
    B-array.}
  \label{fig:model_comp}
\end{figure}

The target MRF was assumed to be equal to RMS$_{\rm V}\times f_{\rm
  MRF}$, where $f_{\rm MRF}$ is a  constant multiplicative factor. The
canonical value for $f_{\rm MRF}$ was determined in two ways. Firstly,
artificial point sources were injected into the {\it uv}-data for an
emission-free field and the {\scriptsize OBIT} {\scriptsize Imager}
task was applied using a range of values for $f_{\rm MRF}$. After each
iteration the recovered 
flux and RMS noise of the final image was measured. 
Figure~\ref{fig:clnflux_vs_mult} shows the results of this
experiment. We found in practise that values of $f_{\rm
  MRF}=0.8\,-\,2.0$ recovered greater than 96 percent of the flux,
within the errors. Secondly, we chose a representative sample of
compact sources with simple morphologies and inspected the
deconvolved images for residual sidelobe structure. Values of $f_{\rm
  MRF}=0.8\,-\,1.0$ were required to fully remove sidelobe structure
from the images. We re-imaged fields using a higher threshold where
obvious clean artifacts were present. Figure~\ref{fig:fmrf} shows the
values of $f_{\rm MRF}$ used across the survey. Most of the $\crn$
area was imaged 
using an $f_{\rm MRF}=0.8$. Fields at lower declinations (including
the BnA observations) required higher $f_{\rm MRF}$ values to avoid
producing increased numbers of low-level artefacts in otherwise
empty regions. Unresolved weak detections are dominated by the
extragalactic population and hence have a flat distribution on the 
sky (\citealt{Anglada1998}, see also Sections~\ref{sec:spurious}
and~\ref{sec:src_gal_distrib}). Based on an initial pass at the  
data reduction we used two further levels of $f_{\rm MRF}=1.0$
and~1.1, chosen to keep the number of 5\,--\,6$\sigma$ point-sources
roughly constant away from the Galactic mid-plane. However, multiplier
values between 0.8 and 1.1 lead to imaging artefacts in a minority of
fields affected by poor calibration, containing very bright point
sources ($>$1\,Jy) or extended emission. Such fields were inspected
and cleaned manually. A small proportion of fields ($\ll1$\,percent)
were found to contain significant circularly polarised flux and were
also cleaned by hand. It is likely that this emission is due to the
instrumental beam squint, rather than real emission on the sky. The
{\scriptsize OBIT} {\scriptsize Imager} task does not correct for beam
squint, however, tests on selected $\crn$ data did not find any
believable Stokes~{\it V} after a correction had been applied
(B.~Cotton, private 
communication). The manually cleaned fields are mostly confined to
high-mass star-formation regions. They appear on Figure~\ref{fig:fmrf}
as patches of red hexagons clustered around the mid-plane of the
Galaxy. Values used in these cases ranged over $1.2\le f_{\rm 
MRF}\le10$, with a mean of 3.2.

Applying the deconvolution algorithm close to the noise can result in
an increase in the so called `clean bias', an effect which results in
a systematic reduction in object fluxes. It is believed to be caused
by inadvertently {\scriptsize CLEAN}ing bright sidelobes, leading to a
subtraction of real flux from astronomical objects \citep{White1997,
  Condon1998}. We measure the clean bias for $\crn$ images in
Section~\ref{sec:clean_bias}.


\subsubsection{Imaging extended emission}\label{sec:img_extended}
Baseline lengths on the VLA B and BnA arrays range from approximately
300\,k$\lambda$ to 2\,k$\lambda$, equivalent to spatial scales of
1.5\,$''$ to 2\,$'$, respectively. These array configurations sample the {\it
  uv}-plane less well at shorter spacings and the deconvolution
algorithm has difficulty reconstructing image structure on scales 
greater than $\sim14''$. 
The imaging procedure tends to produce `waves' or `ruffles' in the
background of fields containing significant extended emission. Flux is
also scattered over the image as the standard {\scriptsize CLEAN}
algorithm attempts to model the emission as a series of delta
functions. This can lead to high RMS noise levels and multiple imaging
artefacts, especially if self-calibration is allowed to run unchecked. 
A total of 193 fields ($\sim2\,\%$) were found to have
poorly-imaged extended emission. To combat this problem we
imaged these fields using a custom weighting scheme. By default the
$\crn$ pipeline is configured to use robust weighting
\citep{Briggs1995}, which is a compromise between the low thermal noise of
natural weighting and the high resolution of uniform weighting. 
In the case of uniform weighting the {\scriptsize OBIT Imager} weights
each visibility by the sum of the weights present in each cell in the {\it
  uv}-plane. For fields with extended 
emission we have instead weighted by the inverse of the number of
visibilities within a radius of ten cells, attenuated by a Gaussian
function. For the B and BnA arrays this has the 
effect of weighting down the poorly sampled short spacings, similar to
the effect of applying an inverse taper. We found that the RMS noise
and number of artefacts in the imaged fields are reduced at the expense of
additional `missing' flux. The reduction in flux compared to a
uniform, or robust weighted image is highly dependent on the structure
of the emission.  

In order to quantify the effect of the two weighting schemes (robust\,=\,0
and Gaussian) we imaged artificial Gaussian sources of increasing size
inserted into the {\it  uv}-data for a blank field. The peak flux 
was fixed at 1.0\,Jy\,beam$^{-1}$ 
while the full-width half-maximum was increased from 2.0$''$ to
26$''$ in steps of 2$''$. Figure~\ref{fig:missing_flux} ({\it top panel})
plots the recovered flux as a function of the source FWHM for both
weighting schemes. When imaging using robust weighting, the fraction
of recovered flux drops off above FWHMs greater than
$\sim8''$. When using the Gaussian-smoothed weighting scheme this
drop-off also occurs at FWHM $\ge8''$, however, the fraction of
recovered flux falls more rapidly. The {\it bottom panel} of
Figure~\ref{fig:missing_flux} shows the RMS noise as a function of
injected source size. In this case the flux density was fixed at
1.0\,Jy to avoid being dynamic-range limited at higher
fluxes. It is clear that the RMS noise in the Gaussian weighted
images is significantly lower and more stable as a function of
FWHM. In practise dynamic ranges of several thousand are achieved on
isolated point sources, falling to several hundred for slightly
resolved sources ($>1.8$\,arcsec).

Figure~\ref{fig:model_comp} shows the effect of the two weighting
schemes on the fidelity of the images. In the top panel is plotted the
absolute fractional value  of the residual flux remaining after the
model image is subtracted from the pipeline imaged data
($|S_{\rm resid}|/S_{\rm model}$). Lower numbers mean that the image is similar to
the model, while higher numbers mean that there are significant structural
differences. It can be seen from the plot that the Gaussian weighting
scheme is the most consistent at representing the source morphology,
while the robust weighted images break down between FWHM\,=\,$8''$
and~$10''$. To further quantify the effect we fit the pipeline imaged
data with a 2D-Gaussian using the {\scriptsize MIRIAD} task
{\scriptsize imfit}. In the bottom panel of
Figure~\ref{fig:model_comp} is plotted the fitted versus injected
FWHM. It is again clear that the robust weighted images begin to
differ from the model at FWHM$\approx10''$, while the 
Gaussian weighting scheme preserves structures out to $\sim14''$. Note
that real-world emission with complex morphology will react
differently to the Gaussian weighting scheme, depending on its
visibility function.


\begin{figure}
  \centering
  \includegraphics[width=8.2cm, trim=0 0 0 0]{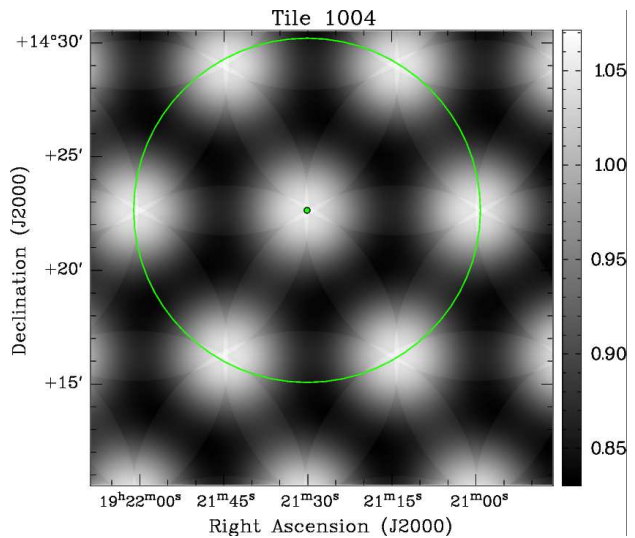}
  \caption{\small Example of a weight image used in the mosaicing
    process. The greyscale depicts the squared beam patterns
    $P^2(\rho)$ accumulated onto a $20'\times20'$ tile. The green
    circle shows the extent of a single field and the dot its pointing
    centre.}
  \label{fig:wt_img}
\end{figure}

\subsection{Mosaicing}\label{sec:mosaicing}
All fields were imaged out to a radius of eight arcminutes ($\sim10$
percent power pattern) before being linearly mosaiced in the image
plane onto $20'\times20'$ tiles oriented in equatorial (J2000)
coordinates. In total 1408 tiles 
cover the survey area, each of which overlap by $1'$. Pointing centres
sit on a close-packed hexagonal grid adapted from the 1.4\,GHz NVSS
survey and scaled to 5\,GHz. \citet{Condon1998} justifies this layout
in detail and \citet{Hoare2012} describes the implementation
in $\crn$. Here we provide a summary for convenience. Adjacent $\crn$
pointing centres are separated by 7.4\,$'$, compared to the 8.9\,$'$
full-width half-maximum of the primary beam at 4.86\,GHz. The
separation is optimised to maximise the uniformity of the noise
pattern without appreciably degrading observing efficiency. At any
point in the mosaic the sky brightness $B$ is given by a weighted sum
of the individual brightness values $b_i$ contributed by the
overlapping snapshots
\begin{equation}
B = \left. \displaystyle\sum\limits_{i=1}^m\,W_i~b_i
\middle/
\displaystyle\sum\limits_{i=1}^m W_i^2. \right.
\end{equation}
To maximise sensitivity the weighting factor $W_i$ was set to be
proportional to $P(\rho)$, the primary beam pattern as
a function of offset $\rho$ from the pointing centre. This correction
is necessary as the noise is constant across a raw snapshot
image and must be weighted by the square of the signal-to-noise-ratio.
The weighting method is implemented in the $\crn$ pipeline in two
steps. Individual fields are first multiplied by $P(\rho)$ and summed onto
a blank tile. This image is then divided by a `weight image'
created from the  
sum of $P^2(\rho)$ functions (modelled by Gaussians for the VLA). The
resultant data product is a mosaiced image which has been primary beam
corrected (i.e., divided by $P(\rho)$). Figure~\ref{fig:wt_img}
illustrates an example of a weight image. The minimum weight is
$P^2(\rho)=0.83$, hence the worst-case relative-sensitivity is
$\sqrt{P^2(\rho)}=0.91$.


\begin{figure*}
  \centering
  \includegraphics[width=18.0cm, trim=0 0 0 0]{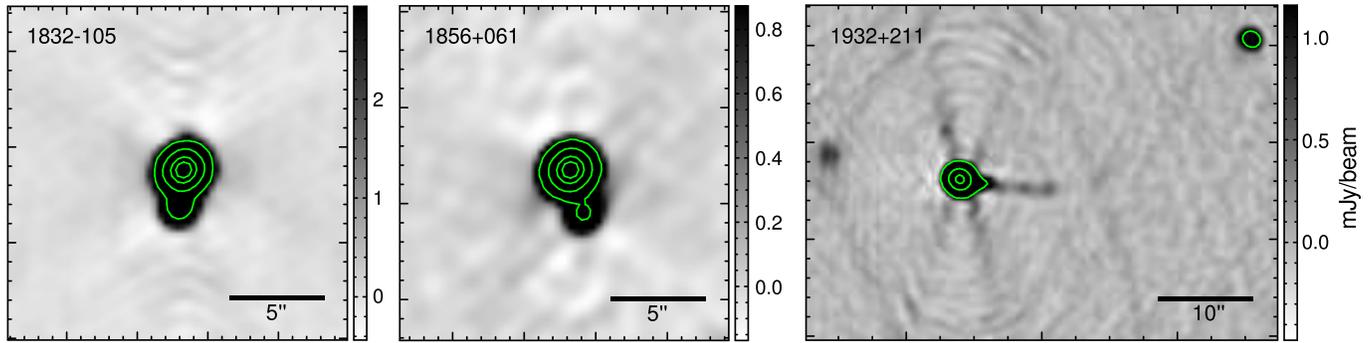}
  \caption{\small Images of the secondary calibrators used to perform phase
    tracking. All three quasars exhibit jets, while 1925+211 has an
    extended jet and there are several sources brighter than 1\,mJy
    within one arcminute. The image scales have been stretched to show all
    real emission, but also highlight very low-level imaging
    artefacts. Only clean-components from real emission were used 
    when calibrating the data.}
  \label{fig:sec_cal_imgs}
\end{figure*}

\section{Data quality}
Data were reduced and imaged for quality control purposes immediately
after the observations were completed. Bad data were quickly identified
allowing the affected fields to be re-scheduled in the observing
queue. The rapid turn-around time meant that we were able to
re-observe most fields affected by poor weather in 2006 and system power
glitches in 2007/2008.


\begin{figure}
  \centering
  \includegraphics[width=8.2cm, trim=10 0 40 30]{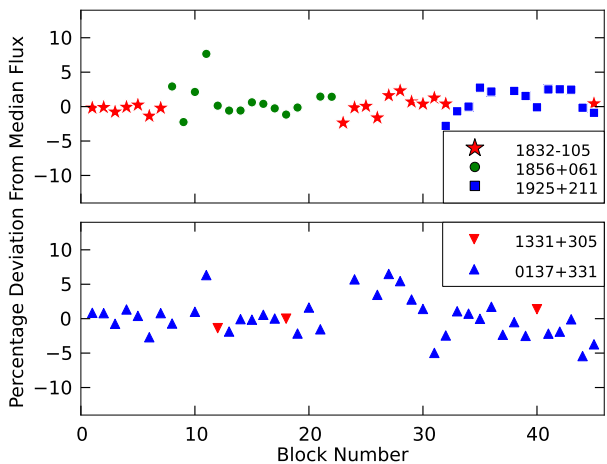}
  \caption{\small {\it Top panel:} Plot of the percentage deviation from
    the median flux density versus block number for the three
    secondary calibrators. Each point represents the integrated flux
    density measured from an eight hour observing block. The scatter
    is mostly within five percent. {\it Bottom panel:} The same plot
    as above except for the backup primary calibrator.} 
  \label{fig:calib_stability}
\end{figure}

\subsection{Calibration}\label{sec:calib_stability}
Three quasars, spaced equally along the plane of the Galaxy, were used
as secondary (phase) calibrators for the whole survey. Although initially
assumed to be point-like, we found that each exhibited structure at
the 0.1 to 2.0 percent 
level, in the form of radio-jets and nearby confusing 
sources. Using the full complement of data available we imaged and
self-calibrated each secondary calibrator field out to the full-width
half-power radius. The resulting clean-component models were used as
inputs to the calibration procedure. Images of the secondary
calibrators are presented in 
Figure~\ref{fig:sec_cal_imgs}. Quasars 1832-105 and 1856+061
deviate from point sources, exhibiting jets with flux densities
peaking at two percent of the main peak. The source 1925+211 shows
significant structure within one arcminute of the central source, including
an elongated jet and two point sources of 1\,mJy and 6\,mJy (0.1
and 0.4 percent of the main peak, respectively).

Two primary flux density calibrators were observed, providing a
redundant means of flux-calibrating the data. 1331+305 (3C286) was
observed at the beginning of an observing block and 0137+331 (3C48) at
the end. With a 5\,GHz flux density of 7.47\,Jy 1331+305 was the preferred
calibrator. However, 0137+331 ($S_{\rm 5\,GHz}=5.48$\,Jy) was used if
technical or weather-related problems affected the initial data from a
block.

The small number of calibrators observed allowed us to check the
consistency of our calibration with time. 
Figure~\ref{fig:calib_stability} shows the percentage deviation from
the median flux densities of the three secondary calibrators and the
backup primary calibrator. Each point on the plot represents an 8-hour block of
observations. Calibrator flux densities were measured directly from
the image data by manually drawing a polygon around each quasar and
summing the flux within the polygon. For the secondary calibrators the
standard-deviation in flux is 2.7~percent, and for the backup
calibrators 8.9~percent, consistent with the accuracy of previous VLA
surveys (e.g., \citealt{Condon1998}, who quote three percent at
1.4\,GHz). No variation with time is seen, implying that the
calibration is stable over the two observing seasons. The scatter in
the backup calibrator is a more appropriate error to quote for
snapshot imaging and is adopted as the formal amplitude calibration
error for $\crn$ data. 

All $\crn$ observations are phase-referenced to one of the three
secondary calibrators and hence adopt their positional
uncertainties. The formal positional uncertainties may be found in the
VLA calibrator manual\footnote{http://www.vla.nrao.edu/astro/calib/manual/} 
and are $<\,150$\,milliarcseconds (mas) for 1832$-$105, $<\,10$\,mas
for 1856+061 and $<\, 2$\,mas for 1925+211.


\subsection{Synthesized Beam shape}
The dual-snapshot observing scheme was designed to deliver the most
circular synthesised beam possible, while allowing both snapshots to
be taken within a single eight-hour observing block. To minimise the
total range of synthesised beam shapes in the survey each field should
ideally be observed at an equal $\pm3$\,hr hour-angle before and after
its zenith position. Scheduling constraints meant that this was not
achieved in practise and a compromise of four hours between snapshot
images was implemented. \citet{Hoare2012} presents the parameters of
the synthesised beams attained in the final images, which we briefly
summarise here.

Within each observing block the beam elongation increases towards
lower declinations, while the position angle varies by
$\sim60$~degrees. The distribution of beam minor-axes in the survey
area separates into two distinct populations, with a small peak at
$0.77''$ and a large peak at $1.2''$. The smaller peak stems 
from the low-declination fields observed using the BnA array
configuration, while the larger one contains the majority of fields
observed using the B array. In contrast, the distribution of major axes
values is monolithic, with a median at $1.5''$ and a standard
deviation of $0.32''$. Ninety-eight percent of fields have elongations
less than two and seventy-four percent less than 1.5. 

\begin{figure}
  \centering
  \includegraphics[angle=-90, width=8.2cm, trim=0 0 0 0]{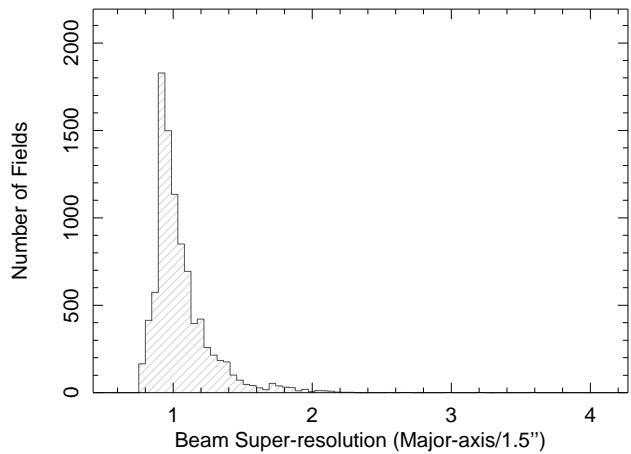}
  \caption{\small Histogram showing the distribution of super-resolution
    caused by forcing a $1.5''$ circular restoring beam for all
    pointing positions.}
  \label{fig:hist_beam_params}
\end{figure}

Based on these values, we chose to force a circular restoring beam of
FWHM 1.5$''$ because this greatly simplified the mosaicing operation
and meant that the restoring beam shape was constant across every
mosaiced image. The value 1.5$''$ was chosen as the median value of
the measured major-axes from all $\crn$ fields. The degree of
super-resolution is presented in Figure~\ref{fig:hist_beam_params} and
is less than 1.5 in ninety-six percent of fields. The restoring beam
{\it area} is larger than the synthesised beam area for 8,154 fields
(87.2 percent) and is less than 1.5 times greater in 9,343 fields
(99.9 percent). 

\begin{figure*}
  \centering
  \includegraphics[width=18.0cm, angle=0, trim=0 0 0 0]{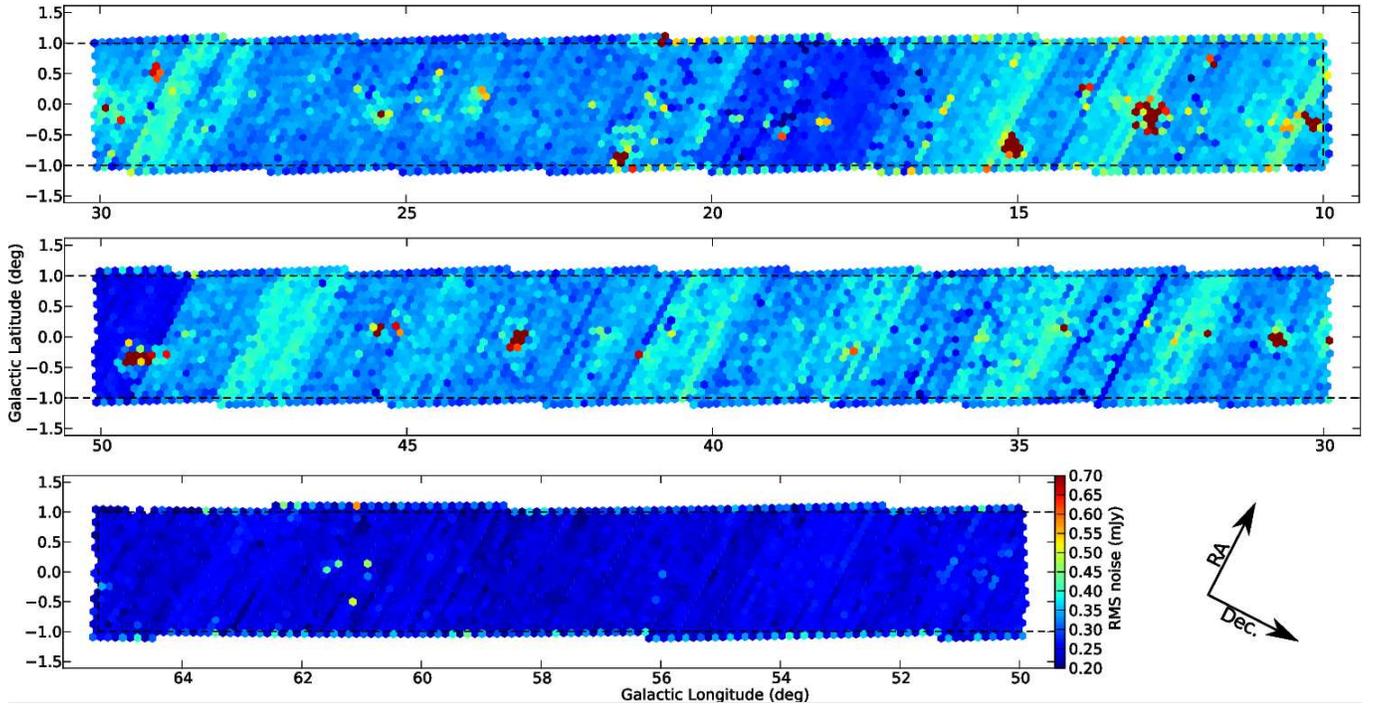}
  \caption{\small Map of the RMS noise in each field of the $\crn$
    survey. Striping in right ascension is due to changes in observing
  conditions between scan-rows. Clusters of high-noise fields (red)
  occur at the locations of star-forming complexes, which contain
  bright and extended emission.} 
  \label{fig:noisemap}
\end{figure*}


\subsection{Sensitivity and uniformity}
\begin{figure}
  \centering
  \includegraphics[height=8.2cm, angle=-90, trim=0 0 0 0]{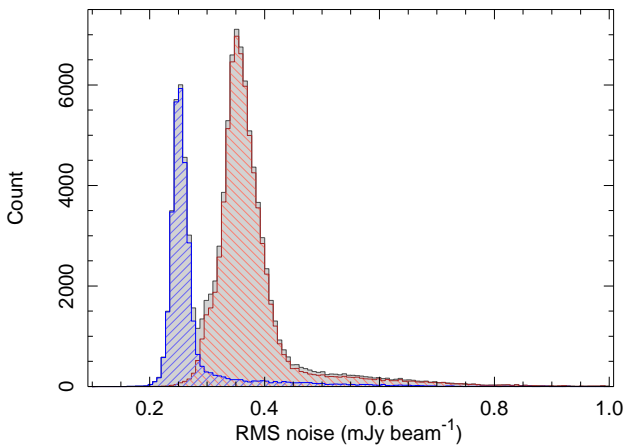}
  \caption{\small Distribution of the total noise over
    the survey region made by sampling on $2'$ scales (solid
    histogram). The right-slanted histogram (blue) contains only data
    from the epoch IIIb observations ($\delta>14.2\degree$) while the
    left-slanted histogram (red) contains the remainder (epochs I, II
    and IIIa). } 
  \label{fig:hist_noise}
\end{figure}

Figure~\ref{fig:noisemap} presents an image of the RMS noise over the
full survey area, with each colour-coded hexagon representing a
field. The locations of $\hii$ region complexes are prominent as
clumps of high-noise fields located close to the 
mid-plane of the Galaxy. Away from such regions the noise level 
within individual scan-rows (scanning in RA) is relatively constant
compared to the variation between rows, which is largely weather
related. The observation area can be divided into two regions with
noticeably different noise properties. At declinations greater than
$\delta=14.2\degree$, the median RMS noise is significantly lower
(RMS$_{\rm outer}$\,=\,0.25\,mJy\,beam$^{-1}$) than the 
remainder of the survey area (RMS$_{\rm inner}$\,=\,0.35\,mJy\,beam$^{-1}$).
This outer $\crn$ region corresponds directly to the epoch-IIIb observations
detailed in Table~\ref{tab:obs}. From the 2007 season onwards the VLA
made extensive use of the upgraded EVLA antennas, which have more
sensitive receivers. In addition, the weather conditions were better
in the second season than in 2006, when observations were affected
by electrical storms. Observations of approximately the inner 20
degrees of the $\crn$ area ($\delta<-10.5\degree$) also took place
during the second season, corresponding to epochs II and
IIIa. However, the RMS noise 
level is similar to the 2006 season for 
a number of reasons. In particular, the inner $\crn$ region is seen
at relatively low elevations from the VLA site, requiring the
telescope to peer through a greater path-length of
atmosphere. Emission from the atmosphere causes an increase in system
temperature decreasing the signal-to-noise ratio in the data. The
epoch-II observations utilised fewer EVLA antennas and, because the
telescope was at the beginning of the VLA/EVLA transition, required
extensive flagging to render the data
usable. Figure~\ref{fig:hist_noise} presents a histogram of the
distribution of noise measurements, sampled on $2'$ scales, across the
whole survey. 
The division between the inner and outer $\crn$ regions is
obvious. Both regions exhibit high-noise tails, corresponding to
fields containing bright and extended emission.

\begin{figure}
  \centering
  \includegraphics[width=8.2cm, angle=0, trim=0 0 10 10]{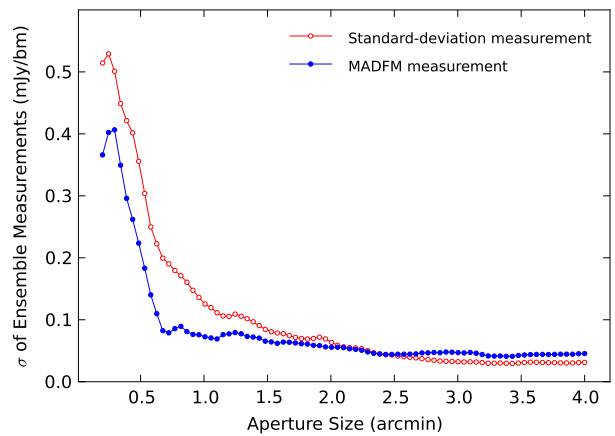}
  \caption{\small Results of repeated noise measurements performed
    using a range of square apertures on $\crn$ data containing a
    ripple. The y-axis plots the scatter in the ensemble set of
    measurements as a function of aperture size. It is clear the
    MADFM values are robust for apertures with scales greater than
    $40''$.}
  \label{fig:noise_scale}
\end{figure}


\subsubsection{Spatial scale of noise}\label{sec:noise_scales}
Interferometry data often exhibit non-Gaussian noise statistics,
largely due to the non-linear deconvolution process and poorly sampled {\it
  uv}-coverage at large spatial scales. In regions with complex
structures on scales greater than $\sim14''$ the emission is poorly
constrained by {\it uv}-coverage of the VLA B arrays. If only a few
short baselines contain most of the flux a simple fringe pattern is
produced on the sky. The flux is not evenly distributed but
accumulates at specific spatial scales, depending on the sampling in
{\it uv}-space. The deconvolution algorithms used here also struggle
to model this emission, resulting in some of the flux being scattered
onto the surrounding sky (see Section~\ref{sec:imaging}). It is
important to characterise this `ripple' noise pattern before
attempting to search for real emission in the $\crn$ data.

We have measured the noise characteristics of representative $\crn$
data affected by a ripple. The region chosen was centred on
$\alpha\,=\,18^h09^m21.96^s,$ $\delta\,=\,-20^{\circ}19'34.9''$ and the
RMS noise was measured using both the standard-deviation (STDEV) and
{\it median absolute deviation from the median} (MADFM)
statistics. For a dataset X = x$_1$, x$_2$ \ldots\,x$_i$ \ldots\,x$_n$
 MADFM is given by   
\begin{equation}\label{eqn:madfm}
  {\rm \sigma = K~median\,(|x_i - median(X)|)},
\end{equation}
i.e., the median of the deviations from the median value. For a normal
distribution MADFM is equivalent to the standard deviation using a scale
factor K\,=\,1.4826. The advantage of MADFM is that
it is insensitive to the presence of outliers in the distribution and
delivers a robust estimate of the true noise. Measurements
were conducted using a range of aperture sizes, varying between $12''$
and $240''$ in steps of $2.82''$. In total twenty one positions were
measured, offset in declination by $6''$ along a line centred on the
noise peak. The scatter in the results (expressed in standard
deviations $\sigma$) for each aperture size is plotted in
Figure~\ref{fig:noise_scale}. From the plot we see that the scatter in
the ensemble set of measurements increases as the aperture size
decreases. The MADFM statistic remains stable at smaller spatial
scales than the STDEV. At scales less than $2'$ the scatter in the
STDEV measurement slowly rises, compared to MADFM, whose scatter
remains less than 0.1\,mJy\,beam$^{-1}$ until scales of
$40''$. Measurements of the global noise-properties of the $\crn$ data
are therefore best performed using apertures spanning $40''$ or larger
using the MADFM statistic.


\section{The $\crn$ source finder}
We have developed an automated source finding procedure with the aim
of producing a well-characterised catalogue of 5\,GHz emission
in the northern Galactic plane. In the following subsections we describe
the source-finding and measurement procedures and investigate the limits of
the catalogue.
 

\begin{figure*}
  \centering
  \includegraphics[width=17.0cm, angle=0, trim=0 0 0 0]{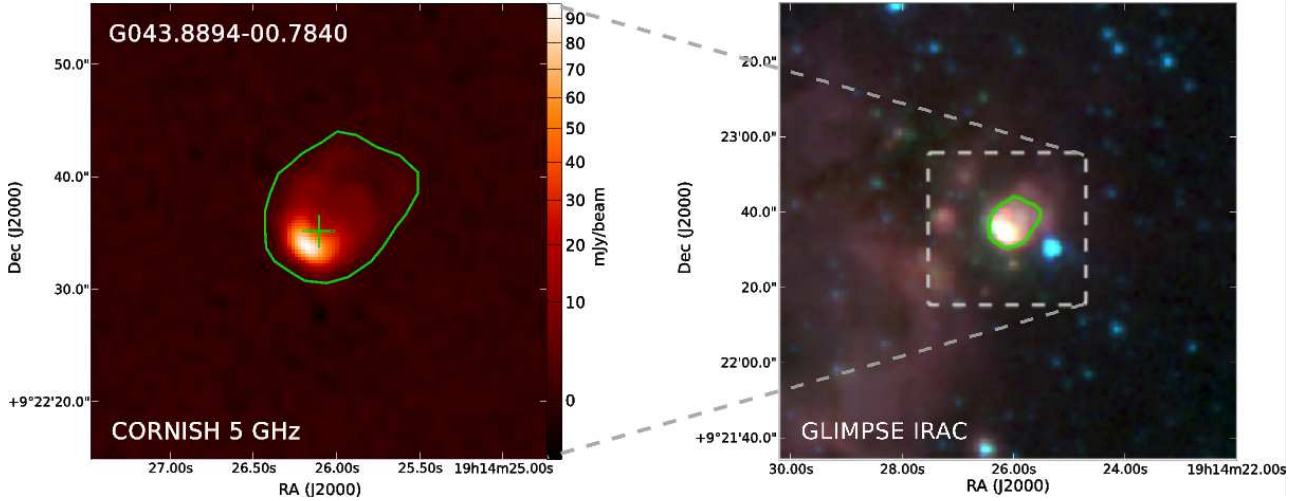}
  \caption{\small {\it Left panel}: Example of a polygonal photometry
    aperture drawn around a spatially extended $\crn$ detection, in
    this case the $\uchii$ region G043.8894$-$00.7840. The
    integrated flux density is calculated from the enclosed pixels
    minus an average background flux measured from the sky in the
    vicinity of the source. A cross marks the intensity-weighted
    position. {\it Right panel:} The $\gli$ 3-colour image
    (red\,=\,8.3\,$\micron$, green\,=\,4.5\,$\micron$, blue\,=\,3.6\,$\micron$)
    exhibits an overall morphology similar to that of the radio
    emission.}
  \label{fig:polygon_example}
\end{figure*}

\subsection{Source detection and photometry}
Tiles were automatically searched for emission using
a custom procedure based on the {\scriptsize OBIT} {\scriptsize FndSou}
task. {\scriptsize FndSou} identifies contiguous islands of emission
above a global intensity threshold and attempts to fit one or more 2D
Gaussians to each. This approach works well in the simplest case of an
image with homogeneous noise properties, however, in the worst-case
scenario the RMS noise can change by a factor of a few over a
$20'\times20'$ tile. This is especially true of tiles covering the
Galactic mid-plane, where massive star-forming complexes are
common. Using a single intensity threshold often results in spurious
detections or omissions of real sources. To compensate for variable
noise levels we ran {\scriptsize FndSou} on a 9$\times$9 grid of
`patches' within the tile area. Each patch is 800 pixels (4\,$'$) on a
side and overlaps adjacent patches by 400 pixels in R.A. and Dec. The
local RMS noise in each patch was  determined using a histogram
analysis clipped at 3$\sigma$ from the median value. With this patch
layout a radio source within a $2'$ band around the tile edge may be
detected in two patches, except at the tile  corners. A source in the
interior may be detected in up to four overlapping patches. The
maximum fitted Gaussian FWHM was constrained to be $\le30''$ in
keeping with the {\it uv}-coverage. Fits within 14$''$ of the patch
edge were deemed invalid, except where a patch abutted a tile
edge. Running this patch-based emission finding procedure results in a
degenerate list of sources with  coincident positions derived from
overlapping patches. A list of unique Gaussian fits to each tile was
produced by searching for duplicates at similar positions
(separation\,$<1''$) and with similar peak amplitudes ($A_{\rm
  min}/A_{\rm max}>0.7$). The Gaussian fit closest to the centre of a
patch was retained. 

Initially, the search was conducted using a 4$\sigma$ local
noise threshold and aperture photometry was performed to
weed out detections with a signal-to-noise ratio $\sigma<5.0$
($\sigma=$\,maximum pixel/RMS-noise). An elliptical aperture was used to
measure the source properties, which extended to the 3$\sigma$
Gaussian major and minor axes ($2.548\,\times\,$FWHM). If the emission
was indeed Gaussian in shape this aperture would encompass 99.7
percent of the emitting flux. The RMS noise and median background
level of the sky were measured from a 20$''$ wide
annulus centred on the source and offset from the measurement aperture
by 5$''$. The annulus width was chosen to sample the local noise
pattern without being influenced by ripples or negative-bowls (see
Section~\ref{sec:noise_scales}). In crowded regions the sky annulus is
likely to contain bright and real sources so the noise was measured
using the robust MADFM statistic. The parameters of the valid Gaussian fits
and photometric measurements were both recorded to the MySQL
database, although the Gaussian fits are preferentially used
in the default $\crn$ catalogue.


\subsection{Resolved emission}\label{sec:measure_resolved_poly}
The source finder determines accurate fluxes for isolated and
unresolved sources but decomposes complex structures into multiple
overlapping Gaussians fits. It is highly desirable to merge these into
a single measurement to avoid over-interpreting the number-counts and
properties of sources in the final catalogue. Clusters of Gaussians 
were identified in the catalogue using a friends-of-friends search:
a Gaussian was associated with a cluster if it was within 12$''$
of any other member. In total, 741 clusters were found and these were
all inspected manually. To distinguish between adjacent but unrelated
sources and over-resolved emission the morphology at 5\,GHz was
compared to that in the {\it Spitzer} $\gli$ mid-infrared images. The
most common extended sources in the images are $\uchii$ regions and
planetary nebulae, each of which have distinctive mid-infrared
signatures. For these types of object the morphology of the 8\,$\micron$ emission
often echos that of the radio continuum \citep{Hoare2007}. If a
cluster of Gaussian fits was found to trace an over-resolved source
then the fitted parameters were replaced with a single measurement
under a polygonal aperture manually drawn around the
emission. Figure~\ref{fig:polygon_example} shows an example of a
polygon carefully drawn around the border of a cometary $\hii$
region. The flux density is calculated from the sum of the pixels
within the source aperture minus the median background level in the
vicinity of the source. In addition to the coordinates of the peak
emission (for which the source is named in $l$ and~$b$) we also record
the geometric and intensity-weighted positions.


\subsection{Measurements and uncertainties}\label{sec:errors}
Below we explain how the properties of the sources were measured and
the uncertainties calculated. The final values are presented in the
$\crn$ catalogue, including the measurement-error and the absolute
uncertainty on each parameter, incorporating the calibration error of
8.9 percent.


\subsubsection{Gaussian fits}
Uncertainties on the Gaussian fits are calculated using the
equations derived by \citet{Condon1997}, summarised here for
convenience. Noise in interferometric data is correlated on the scale
of the synthesised beam FWHM, in this case $\theta_{\rm bm}=1.5''$. The
effective signal-to-noise level $\rho$ of a source with measured peak
amplitude $A_{\rm peak}$ seen against a background of correlated
Gaussian noise is given by 
\begin{equation}\label{eqn:corr_gau_noise}
  \rho^2=\frac{\theta_{\rm M}\theta_{\rm m}}{4\theta_{\rm bm}^2}
\left[1+\left(\frac{\theta_{\rm bm}}{\theta_{\rm M}}\right)^2\right]^{\alpha_{\rm M}}
\left[1+\left(\frac{\theta_{\rm bm}}{\theta_{\rm m}}\right)^2\right]^{\alpha_{\rm m}}
\frac{A_{\rm peak}^2}{\sigma_{\rm sky}^2},
\end{equation}
where $\theta_{\rm M}$ and $\theta_{\rm m}$ are the respective major
and minor fitted axes and $\sigma_{\rm sky}$ is the RMS noise measured
directly from the image. The exponents $\alpha_{\rm M}$ and
$\alpha_{\rm m}$ have been estimated by 
\citet{Condon1997} via Monte-Carlo simulations and are 
$\alpha_{\rm M}=\alpha_{\rm m}=3/2$ for the amplitude and flux density errors,
$\alpha_{\rm M}=5/2,~\alpha_{\rm m}=1/2$ for the error on the major axis, and
$\alpha_{\rm M}=1/2,~\alpha_{\rm m}=5/2$ for the minor axes, position angle and
absolute coordinate errors. On average, the signal-to-noise ratio is
{\it increased} by a factor of 1.4. The positional uncertainties parallel to
the major ($\sigma_{\rm M}$) and minor ($\sigma_{\rm m}$) fitted axes
are given by 
\begin{equation}\label{eqn:err_fit_axes}
\frac{\sigma_{\rm M}^2}{\theta_{\rm M}^2} = \frac{\sigma_{\rm m}^2}{\theta_{\rm m}^2}
\approx \frac{1}{(2~ln\,2)\rho^2},
\end{equation}
using values for $\rho$ calculated from
Equation~\ref{eqn:corr_gau_noise}. When the fit is projected onto
equatorial axes the absolute position errors in right ascension
($\sigma_{\alpha}$) and declination ($\sigma_{\delta}$) become
\begin{equation}\label{eqn:err_fit_equatorial}
\begin{array}{@{}l}
\sigma_{\alpha}^2\approx \epsilon_{\alpha}^2+ \sigma_{\rm M}^2\,{\rm sin^2(P.A.)}
+\sigma_{\rm m}^2\,{\rm cos^2(P.A.)}, \\
\sigma_{\delta}^2\approx \epsilon_{\delta}^2+ \sigma_{\rm M}^2\,{\rm cos^2(P.A.)}
+\sigma_{\rm m}^2\,{\rm sin^2(P.A.)},
\rule{0pt}{5mm}
\end{array}
\end{equation}
where P.A. is the position angle of the fitted major axis east of
north and $\epsilon_{\alpha}=\epsilon_{\delta}\approx0.1\,{\rm
 arcseconds}$ is the systematic positional uncertainty. This value was
determined via a comparison between $\crn$ and catalogues of quasars
whose positions are determined to milliarscsecond accuracy. The 15
matching quasars were drawn from the Goddard VLBI astrometric
catalogues\footnote{http://gemini.gsfc.nasa.gov/solutions/}, Very Long
Baseline Array Galactic Plane Survey (VGaPS, \citealt{Petrov2011}) and
the VLA-calibrator
manual\footnote{http://www.vla.nrao.edu/astro/calib/manual/} and their
median offset of 0.1\,arcseconds was adoped as the systematic
positional uncertainty for $\crn$.

Errors in the P.A. may be calculated from 
\begin{equation} 
  \sigma_{\rm
    PA}^2=\frac{4}{\rho^2}\left(\frac{\theta_{\rm M}\theta_{\rm m}}{\theta_{\rm M}^2-\theta_{\rm m}^2}\right)^2,
\end{equation} 
although we note that position angle values are only relevant when
one or more axes is significantly resolved. Uncertainties associated
with the fitted major and minor Gaussian FWHM are given by
\begin{equation}
  \sigma^2(\theta_{\rm M})=\frac{2\theta_{\rm M}^2}{\rho^2} + \epsilon_{\theta}^2\theta_{\rm M}^2
~~~~~{\rm and}~~~~~
\sigma^2(\theta_{\rm m})=\frac{2\theta_{\rm m}^2}{\rho^2}+ \epsilon_{\theta}^2\theta_{\rm m}^2.
\end{equation}
The fractional calibration uncertainty $\epsilon_{\theta}=0.02$ is adopted from the
VLA NVSS survey \citep{Condon1998}, which was observed using a similar
snapshot mode. A single characteristic measured angular size
$\theta_{\rm f}$ may be obtained from the geometric mean of the major
and minor axes 
\begin{equation}
\theta_{\rm f} = \sqrt{\theta_{\rm M}\,\theta_{\rm m}},
\end{equation}
and its associated measurement uncertainty given by
\begin{equation}
\sigma(\theta_{\rm f}) =
\frac{\theta_{\rm f}}{2}\sqrt{\frac{\sigma^2(\theta_{\rm M})}{\theta^2_{\rm M}} +
\frac{\sigma^2(\theta_{\rm m})}{\theta^2_{\rm m}}}.
\end{equation}
The restoring beam was forced to be a circular Gaussian of FWHM 
$\theta_{\rm bm}=1.5''$ over the whole survey area so the
deconvolved source size $\theta_{\rm s}$ in arcseconds may be found from 
\begin{equation}
\theta_{\rm s} = \sqrt{\theta_{\rm f}^2 - 1.5^2},
\end{equation}
although we note that detections with $\theta_{\rm f} < 1.8''$ are considered
unresolved in $\crn$. 

The fitted amplitude $A_{\rm peak}$ must be corrected for the clean
bias $\Delta A_{\rm cb}\,=\,-\,0.94\,\sigma_{\rm sky}$, which we
measured in Section~\ref{sec:clean_bias} below, so
\begin{equation}
  A = A_{\rm peak} - \Delta A_{\rm cb}.
\end{equation}
The uncertainty on the fitted amplitude may be calculated
from 
\begin{equation}
  \sigma_A^2=\frac{2A^2}{\rho^2} + \epsilon_A^2 A^2,
\end{equation}
where $\epsilon_A\,=\,0.089$ is the fractional amplitude calibration error.
Finally, the integrated flux density $S$ under a 2D Gaussian is given by
\begin{equation}
  S =\frac{A\,\pi}{4\,ln(2)}\,\frac{\theta_{\rm M}\,\theta_{\rm m}}{\theta_{\rm bm}^2}
\end{equation}
and the corresponding uncertainty is
\begin{equation}
\frac{\sigma_S^2}{S^2}\approx\frac{\sigma_A^2}{A^2}+\frac{\theta_{\rm bm}^2}{\theta_{\rm M}\theta_{\rm m}}\,
\left[\frac{\sigma^2(\theta_{\rm M})}{\theta_{\rm M}^2}+\frac{\sigma^2(\theta_{\rm m})}{\theta_{\rm m}^2}\right].
\end{equation}


\subsubsection{Aperture photometry}
The peak amplitude reported for sources measured using aperture photometry
is simply the intensity of the brightest pixel within the source
aperture, corrected for the clean bias.
\begin{equation}
  A = A_{\rm max} -\Delta A_{cb}.
\end{equation}
The effective signal-to-noise of the source in the presence
of correlated Gaussian noise may be determined from a modified version of
Equation~\ref{eqn:corr_gau_noise} in which $\theta_{\rm M}$ and
$\theta_{\rm m}$ are both replaced with the intensity-weighted diameter
\begin{equation}\label{eqn:rad_in_wt}
\theta_{\rm d} = 2 \left. \displaystyle\sum\limits_{i=1}^{N_{\rm src}} r_i A_i
\middle/
\displaystyle\sum\limits_{i=1}^{N_{\rm src}} A_i. \right.
\end{equation}
For a perfectly circular Gaussian source
$\theta_{\rm d}\,\equiv\,\theta_{\rm f}\,\equiv\,{\rm FWHM}$.
In Equation~\ref{eqn:rad_in_wt} $r_i$ is the angular distance from the
$i^{\rm th}$ pixel to the brightness-weighted centre and $A_i$ is its
intensity. The sky-noise corrected for Gaussian correlation is then 
\begin{equation}
  \frac{1}{\sigma_{\rm g}^2} = \frac{\theta_{\rm d}^2}{4\theta_{\rm bm}^2}
\left[1+\left(\frac{\theta_{\rm bm}}{\theta_{\rm d}}\right)^2\right]^{\alpha}\frac{1}{\sigma_{\rm sky}^2},
\end{equation}
where $\alpha=3$ for all errors. The uncertainty on the peak amplitude
is
\begin{equation}
  \sigma_A^2 = \sigma_{\rm g}^2 + \epsilon_A^2 A_{\rm max}^2,
\end{equation}
where $\epsilon_A\,=0.089$ is the calibration error.

The equation for measuring the integrated flux density $S_{\rm
  phot}$ of a source using aperture photometry can be written as
\begin{equation}
S_{\rm phot}=\left(\sum\limits_{i=1}^{N_{\rm src}} A_i - N_{\rm
  src}\,\bar{B}\middle)\right/ a_{\rm bm},
\end{equation}
where $\sum_{i=1}^{N_{\rm src}}A_i$ is the total flux in the source
aperture summed over $N_{\rm src}$ pixel elements (in units of
Jy\,pixel$^{-1}$), $\bar{B}$ is the background flux density estimated
from the median level in the sky-annulus and $a_{\rm
  bm}=\pi\,\theta_{\rm bm}^2/(4\,ln(2)\,\Delta_{\rm pix}^2)$ is the
beam-area in pixels (28.33\,pixels for $\crn$ data). $S_{\rm phot}$
must also be corrected for clean-bias. If the source is unresolved the
missing flux $\Delta S_{\rm cb}$ is given by 
\begin{equation}
\Delta S_{\rm cb} = -0.94\,\sigma_{\rm sky}\,\frac{\pi}{4\,ln(2)}\,\frac{\theta_{\rm
    bm}^2}{a_{\rm bm}},
\end{equation}
however, because the clean-bias reduces the flux in all clean-components (CCs)
by a constant factor the effect on extended emission is difficult to 
gauge. The minimum number of CCs required to model an extended source
can be estimated from the number of beam-areas $n_{\rm beams}$
subtended by the emission. The integrated flux density is then
\begin{equation}
  S = S_{\rm phot} - \Delta S_{\rm cb} * n_{\rm beams}.
\end{equation}
The error on the integrated flux density may be found from
\begin{equation}\label{eqn:err_app_flux}
\sigma_S^2 =  \left.\left(\sigma({\textstyle \sum} A_i)^2 + 
\frac{\pi\,N_{\rm src}^2\,\sigma_{\rm g}^2}{2\,N_{\rm sky}}\right)\middle/ a_{\rm bm}^2\right.,
\end{equation}
where $\sigma_{\rm g}^2$ is the corrected variance, $N_{\rm src}$ and
$N_{\rm sky}$ are the the number of pixels in the source- and
sky-apertures, respectively. The term $\sigma({\textstyle \sum} A_i)$
is the uncertainty on the sum over the pixels in the source aperture
given by
\begin{equation}
  \sigma({\textstyle \sum} A_i) = \sum\limits_{i=1}^{N_{\rm src}} 
\left(\sqrt{\sigma_{\rm g}^2 + \epsilon_A^2 A_i^2 }~\right).
\end{equation}

The uncertainty on the intensity weighted diameter maybe found from
\begin{equation}
\frac{\sigma^2(\theta_{\rm d})}{\theta_{\rm d}^2} = 
\frac{\sigma_{\rm g}^2\,\sum r_i^2}{(\sum r_i\,A_i )^2} +
\frac{N_{\rm src}\,\sigma_{\rm g}^2}{(\sum A_i)^2},
\end{equation}
where the sums are taken over the pixels in the source-aperture.
For emission measured using a polygonal aperture
the intensity-weighted position is given by
\begin{equation}\label{eqn:pos_in_wt}
\bar{x} = \left. \displaystyle\sum\limits_{i=1}^{N_{\rm src}} x_i A_i
\middle/
\displaystyle\sum\limits_{i=1}^{N_{\rm src}} A_i. \right.,
\end{equation}
where $x_i$ is the right-ascension ($\alpha$) or declination
($\delta$) in the $i^{\rm th}$ pixel. The corresponding error in
$\bar{x}$ is given by 
\begin{equation}\label{eqn:err_pos_in_wt}
\sigma x^2 =   
\left[\,\frac{\sigma_{\rm g}^2\,\sum x_i^2}{(\sum x_i\,A_i )^2} +
\frac{N_{\rm src}\,\sigma_{\rm g}^2}{(\sum A_i)^2}\,\right]\,\bar{x}^2
+ \epsilon_x^2,
\end{equation}
with $\epsilon_x\,=\,\epsilon_{\alpha}\,=\,\epsilon_{\delta}$, the
absolute positional error of the associated phase calibrator.


\begin{figure}
  \centering
  \includegraphics[height=8.2cm, angle=-90, trim=0 0 0 0, clip]{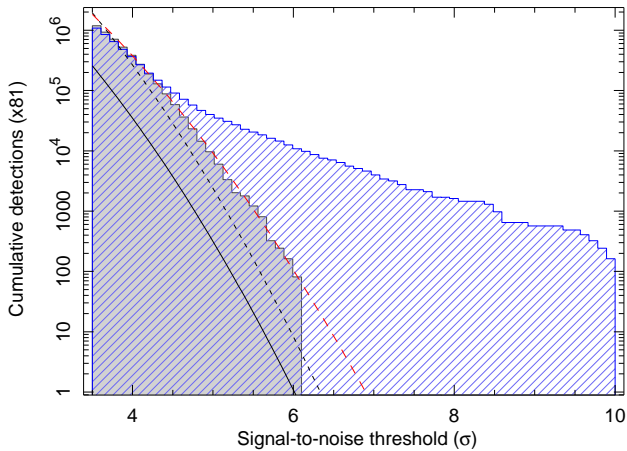}
  \caption{\small The solid histogram illustrates the cumulative
    distribution of spurious detections in the $\crn$ catalogue as a
    function of the signal-to-noise ratio measured from fourteen
    inverted tiles. The plot has been scaled to the total $\crn$ area
    ($\sim1300$ tile areas) by multiplying by 81. In
    comparison, the hatched histogram shows the distribution of
    detections from the same tiles before inversion. See
    Section~\ref{sec:spurious} for further details.}
  \label{fig:hist_spurious}
\end{figure}

\begin{figure*}
  \centering
  \includegraphics[width=18.0cm, angle=0, trim=0 0 0 0]{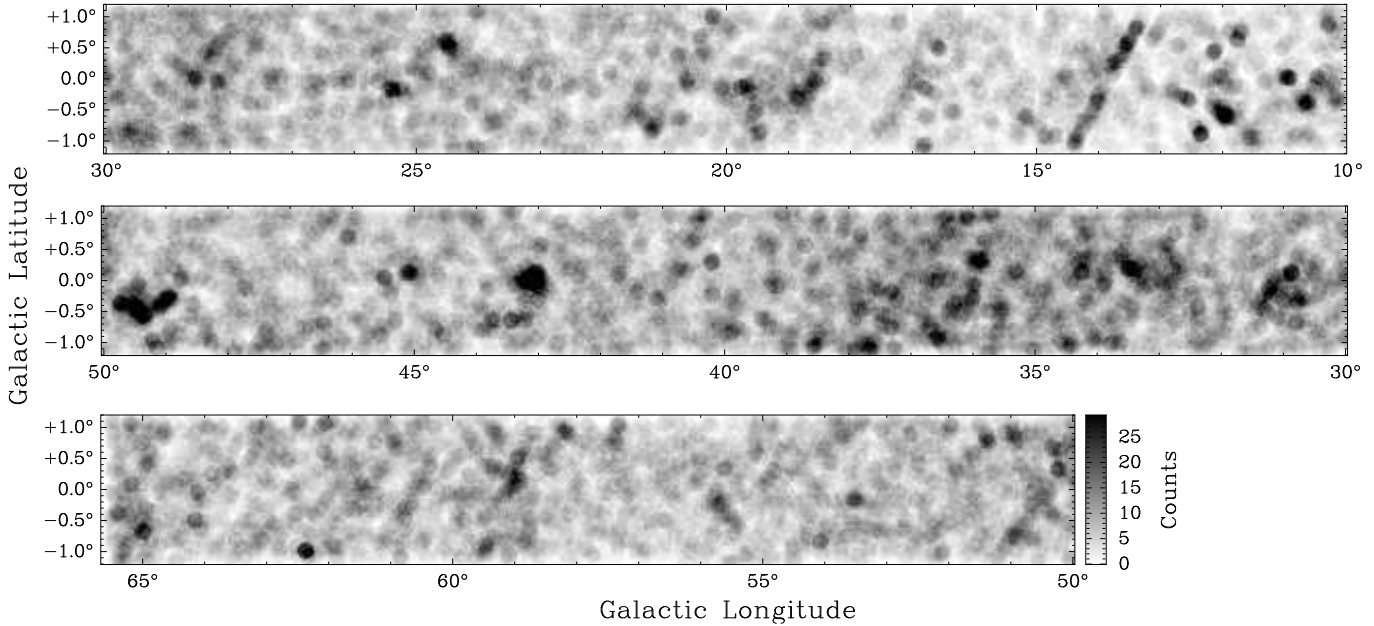}
  \caption{\small Density map of all $\crn$ sources detected above
    $5\sigma$. The greyscale level illustrates the number of sources
    found within an 8 arcmin radius of any position. Away from
    complexes of $\hii$ regions, which are mostly near the Galactic
    mid-plane, variations in the source counts are due to spurious
    sources, e.g., the bad scan-row at $l\,\approx14\degree$.}  
\label{fig:spurious_src_map} 
\end{figure*}

\subsection{Spurious sources}\label{sec:spurious}
We have attempted to estimate the number of spurious sources detected
in {\it well calibrated and well behaved data} by running the source
finder on inverted tiles, i.e., tiles where the pixel values have been
multiplied by $-1$. Any negative detections will be false and allow
us to estimate 
the number of spurious sources as a function of the
signal-to-noise ratio. Fourteen tiles were selected to be  representative of
the emission properties across the survey region. They contained
variously: no strong emission, one or more point sources with
$S_{5\,GHz}>50$\,mJy, weak extended emission, and bright extended sources
causing moderately elevated noise levels
($0.5$\,mJy$<$\,RMS\,$<$0.8\,mJy). For comparison, we ran the source
finder using a $3.5\sigma$ cutoff on both the inverted and the regular tiles. 
Figure~\ref{fig:hist_spurious} plots the cumulative counts of
detections as above a signal-to-noise ratio, expressed as
$\sigma$. The grey-shaded histogram illustrates the number of spurious
detections in the inverted tiles, while the hatched histogram illustrates the
detections in the normal tiles, some of which will be real. Note that
below $4.5\sigma$ the detections are dominated by spurious
sources. The fourteen tiles represent 1.23 percent of the survey area,
so by scaling the plot by 81 we can estimate the total distribution of
spurious sources in $\crn$. 
In Figure~\ref{fig:hist_spurious} the number of spurious sources found
decreases to 81 at 6.1$\sigma$, above which our scaling is too crude
to sample. For populations governed by Gaussian statistics the fraction
$f(\sigma)$ of the populations lying within a $\sigma$-threshold is
given by
\begin{equation}\label{eqn:gauss_erf}
  f(\sigma) = 1 - erf(\sigma/\sqrt2),
\end{equation}
where $erf(\sigma)$ is the Gaussian error function. The solid black curve in
Figure~\ref{fig:hist_spurious} plots $f(\sigma)$
assuming the total number of possible detections is equal to 
the number of synthesised beam areas in 
$\crn$ ($\approx5.6\times10^8$ beams). It is clear that this assumption
underestimates the number of spurious sources found, so we fit $f(\sigma)$
with the total number of sources as a free-parameter. The fit is shown
by the short-dashed line and is dominated by the large number of
sources in the bins with $\sigma<4$. Above
$4\sigma$ the distribution has a shallower fall-off than expected
for Gaussian statistics. An alternative is shown by the long-dashed
line, which is fitted only to the bins with $\sigma>4$ and uses the
error-function of a distribution with a narrower width than purely
Gaussian ($\sigma=0.9\sigma_{\rm gauss}$). It is a significantly
better match to the high signal-to-noise end of the
distribution and predicts less than one spurious source above
$7\sigma$. Based on this reasoning we have chosen a signal-to-noise
cutoff of $7\sigma$ for a high-reliability $\crn$ catalogue. We
caution that data with greater complexity or poor calibration may
introduce significant numbers of false sources above this level, so
this does not mitigate the need to manually inspect the data for
artefacts. Sources detected below $7\sigma$ are not offered as an
official data-product, but this low-reliability catalogue will be made
available on the $\crn$ web page


\subsubsection{Density of low signal-to-noise sources}
A density map of the $\crn$ detections serves to highlight regions
containing excessive numbers of weak sources, some of which may be
spurious. Figure~\ref{fig:spurious_src_map} plots the 
number density of $\crn$ sources above $5\sigma$ summed within an
eight arcminute radius. The most prominent feature is a line of elevated
pixels corresponding to the scan row at $\delta=-16\degree55'03.13''$
($13.3\degree<l<14.5\degree$). This row is unique in the survey as  
each field has only a single 40 second snapshot observation. Data in
the second pass were found to be corrupted and no repeat observations
were scheduled due to an array configuration change.

Isolated clumps of pixels with high source counts (e.g., at
$l=12,$ $l=31$, $l=43$ and $l=49$) correspond to molecular cloud
complexes forming massive stars. The over-density of weak sources in
these regions could be due to either a real increase in source counts
or to an increase spurious in sources generated by the deconvolution
process. Both scenarios warrant careful inspection of the data.


\subsection{Completeness}\label{sec:completeness}
\begin{figure}
  \centering
  \includegraphics[width=8.5cm, angle=0, trim=10 0 50 40]{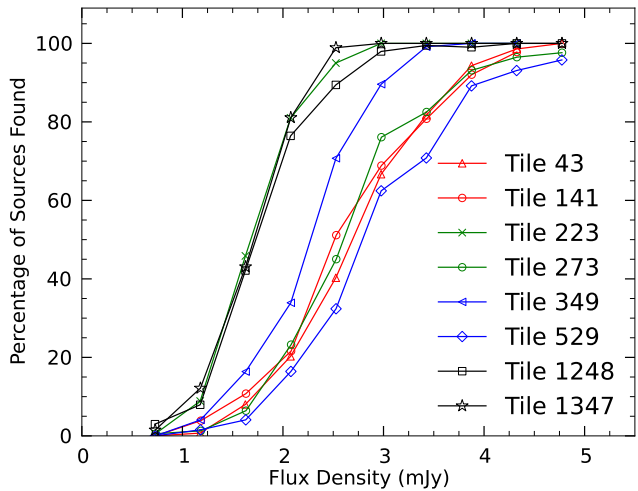}
  \caption{\small Percentage completeness as a function of flux density for
    point sources within eight representative tiles. See the text in
    Section~\ref{sec:completeness} for further details.}
  \label{fig:completeness}
\end{figure}
To quantify the formal sensitivity limits of the pipeline reduced data
we conducted completeness tests on tiles from `empty' parts of the
sky. The tiles were chosen to have few detections above $5\sigma$,
homogeneous noise properties and be free of obvious imaging
artefacts. One-hundred artificial point sources were injected into the 
calibrated {\it uv}-data for each tile before creating a mosaiced image.
The flux densities of the injected sources were varied
randomly between 0.5 and 5.0\,mJy, so as to bracket the expected
sensitivity limit. Positions were also chosen randomly,
but avoided known emission, the tile edges or regions where noise
spikes were common. {\scriptsize FndSou} was then used to find and fit
the emission with Gaussians. After twenty iterations of the
injection-imaging-fitting routine the aggregate results were compared
to the injected source parameters. 

\begin{table}
\caption{Completeness limits of tiles spanning a representative
  range of noise levels.\label{tab:tiles_comp}}
\begin{tabular}{llllll}
\tableline 
\tableline 
Tile & RMS      & $f_{\rm MRF}$ & Epoch & 50\,\% & 90\,\%\\
     & (mJy/bm) &               &       & (mJy)  & (mJy) \\
\tableline                                                    
  43 & 0.384    & 1.1          & II    & 2.7    & 3.8   \\
 141 & 0.384    & 1.1          & II    & 2.5    & 3.8   \\
 223 & 0.299    & 1.0          & IIIa  & 1.7    & 2.3   \\
 273 & 0.358    & 1.0          & IIIa  & 2.6    & 3.8   \\
 349 & 0.344    & 0.8          & I     & 2.2    & 3.0   \\
 529 & 0.367    & 0.8          & I     & 2.8    & 3.9   \\
1248 & 0.239    & 0.8          & IIIb  & 1.7    & 2.5   \\
1347 & 0.257    & 0.8          & IIIb  & 1.7    & 2.3   \\
\tableline
\end{tabular}
\end{table}

Figure~\ref{fig:completeness} plots the percentage of sources
recovered as a function of the flux density for tiles covering a
range of RMS values and Epochs. The image parameters are presented in
Table~\ref{tab:tiles_comp} alongside the fifty and ninety percent completeness
limits. As expected, tiles with lower measured RMS noise levels tend to 
have lower completeness limits. There is, however, significant
variation between tiles as the local completeness limit ultimately
depends on the uniformity of the noise pattern within the tile. At
worst (tile 529, Epoch~I) the $\crn$ survey is 90 percent complete to
point sources at the 3.9\,mJy level.


\subsection{Clean bias}\label{sec:clean_bias}
When deconvolving the synthesised beam from the images, the flux level
at which the {\scriptsize CLEAN} algorithm halts must be chosen
carefully. If the cutoff is set too far above the noise then the
residual images will be dominated by sidelobe patterns. If it is too
low {\scriptsize CLEAN} will inadvertently identify noise-spikes and
sidelobes as real emission. Both negative and positive clean
components on sidelobes will result in flux being subtracted from the
positions of real sources and can artificially lower the RMS
noise. The {\scriptsize AutoWindow} function described in
Section~\ref{sec:imaging} has been shown to reduce the clean
bias. However, we chose a relatively low {\scriptsize CLEAN}
cutoff when imaging $\crn$ data and need to measure the bias level in
order to evaluate the correct flux-densities and uncertainties for the
$\crn$ catalogue. In a 
similar test to the one presented in Section~\ref{sec:completeness} we
inserted twenty point sources into the {\it uv}-data for each of tiles
43, 273, 529 and 1248. The flux densities were set randomly between
2 and 20\,mJy. The data were imaged and mosaiced using the $\crn$
pipeline, and aperture photometry was used to recover the artificial
source fluxes. We found that the clean bias was consistent across all
four tiles, despite being representative of different epochs. We adopt
a mean clean bias of 
$\Delta A_{\rm peak}=-0.94\,\sigma_{\rm sky}$ (typically 0.33\,mJy), indicating a
moderate level of 
over-cleaning compared to NVSS, which quotes $\Delta
A_{\rm peak}=-0.67\sigma$. We judge that this will not affect the
utility of the catalogue. 


\subsection{Manual quality control}
According to the results of Section~\ref{sec:spurious} we do not
expect to find significant numbers of spurious sources above
$7\sigma$ in well-behaved $\crn$ data. This statement is not
necessarily true of high-noise fields containing bright and extended
emission associated with massive star-forming complexes. Occasionally,
peaks in the rippled noise pattern may be mistaken for real emission,
or calibration errors may conspire to create false sources. To
alleviate this problem we visually inspected all
high-reliability $\crn$ detections (i.e., those peaking above
$7\sigma$) to assess them as potential artefacts.

The $\crn$ team visually inspected all mosaic tiles and individual
sources in the $7\sigma$ catalogue. All sources were classified as
being either `unlikely', `possibly' or `likely' an artefact 
based on the criteria above, i.e., located on a peak in a high noise
ripple region, near a very bright source, or in a region where there
appears to be an excessive number of potentially spurious
5\,-\,7$\sigma$ sources (see Figure~\ref{fig:spurious_src_map}). If
a source suspected of being possibly or likely an artefact was
found to have a radio or infrared counterpart then, of course, the flag was
left as `unlikely' in the $\crn$ database. Smaller $\uchii$ regions
lying within the noise radius of a much brighter emission often have
counterparts in the $\gli$ IRAC data, while planetary nebulae (PN) are
often seen in 
the UKIDSS data, confirming them as real detections. Both source types
appear in the far-infrared $\mip$ bands (24\,$\micron$ and
70\,$\micron$). Real extragalactic sources are not likely have
counterparts in the infrared datasets and so retain their
possible-artefact flag in suspect regions.


\section{Results}
We found 3,062 sources in the $\crn$ data above a $7\sigma$
detection threshold. Of these, 2,591 were well fit by model Gaussians
and the remaining 471 sources required measurement using a hand-drawn
polygonal aperture. A total of 286 and 138 sources were classified as
`possible' or `likely' artefacts, respectively, and a flag set in the
final high-reliability catalogue. They remain available in the on-line
catalogue and users will be able to include possible and likely
artefact sources in their searches. Below we present the new,
high-reliability catalogue of 5\,GHz radio-emission containing 2,638 sources.

\subsection{Catalogue format}
Isolated and unresolved sources identified by the source finder 
have two recorded entries taken from fitted Gaussian parameters and
aperture photometry measurements. Sources exhibiting structured and
extended emission have a single entry, based on aperture photometry
performed using a manually drawn polygon. When assembling an aggregate
catalogue we favoured the Gaussian fitted values. The photometric
measurements are useful for diagnostic purposes.

An excerpt from the final $\crn$ catalogue is presented in
Table~\ref{tab:cat7sigma}. The columns are as follows: column (1)
contains the $\crn$ source name, constructed from the Galactic
longitude and latitude of the source. The equivalent right-ascension
($\alpha$) and declination ($\delta$) are displayed in columns (2) and
(3), respectively. For sources well fitted by Gaussians the adopted
coordinates are simply the peak positions of the fits. The intensity
weighted position is quoted for extended sources measured using a
polygonal aperture. The associated positional uncertainties are given
in columns (4) and (5). Two uncertainty values are quoted for
catalogue entries. The first value is the absolute uncertainty,
incorporating both measurement and calibration errors. The second value (in
brackets) is the error associated with the photometry or Gaussian fit
alone. Column (6) presents the peak flux density 
in units of mJy\,beam$^{-1}$. The 5\,GHz integrated flux
density ($S_{\rm 5\,GHz}$) is presented in column (7). Column (8) 
contains the measured angular-scale of the emission $\theta_{\rm f}$,
which has been determined from the geometric average of the major and
minor Gaussian fit axes, or the intensity-weighted diameter in the
case of extended emission. Sources with $\theta_{\rm 
  s}>1.8$ are considered to be resolved in the $\crn$ images and their
deconvolved sizes are presented in column (9). The local RMS noise
measured from the photometric sky-annulus is recorded in column
(10). Column (11) notes how the flux density was measured, either with
an polygonal aperture, or using the Gaussian fit. Finally, column (12)
contains a range of flags notifying the reader if the source is: 
\begin{itemize}
  \item within $12''$ of another source,
  \item lying on an unusually high-noise region (RMS$>0.45$\,mJy\,beam$^{-1}$),
  \item imaged using the smoothed weighting scheme described in
    Section~\ref{sec:img_extended} (within 4.45\,$'$ of a field centre,
    i.e., half a primary beam FWHM), 
  \item within $3'$ of a bright ($>0.5$\,Jy) source,
  \item within $2'$ of the edge of the survey,
  \item within an area containing numerous low signal-to-noise
    detections likely to be spurious,
  \item overlaps with another source,
  \item or has been flagged as a suspected artefact during manual
    inspection.

\end{itemize}
The flags are described in more detail in the footnotes to
Table~\ref{tab:cat7sigma}.

\begin{figure}
  \centering
  \includegraphics[height=8.2cm, angle=-90, trim=0 0 0 0]{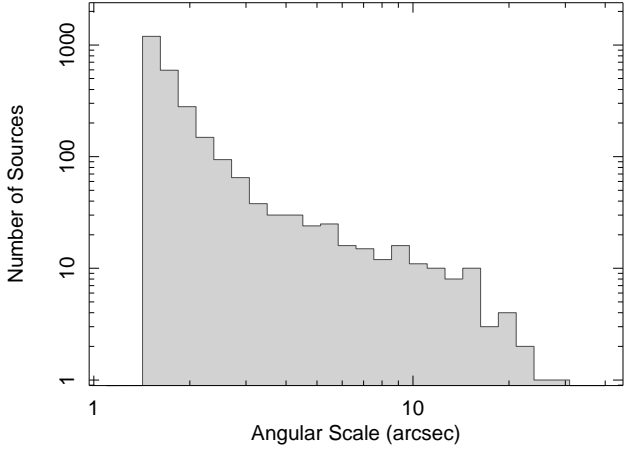}
  \caption{\small Distribution of source angular sizes for the
    high-reliability catalogue.}
  \label{fig:hist_angscale}
\end{figure}

\begin{figure}
  \centering
  \begin{minipage}{80mm}
  \includegraphics[height=7.9cm, angle=-90, trim=0 0 0 0]{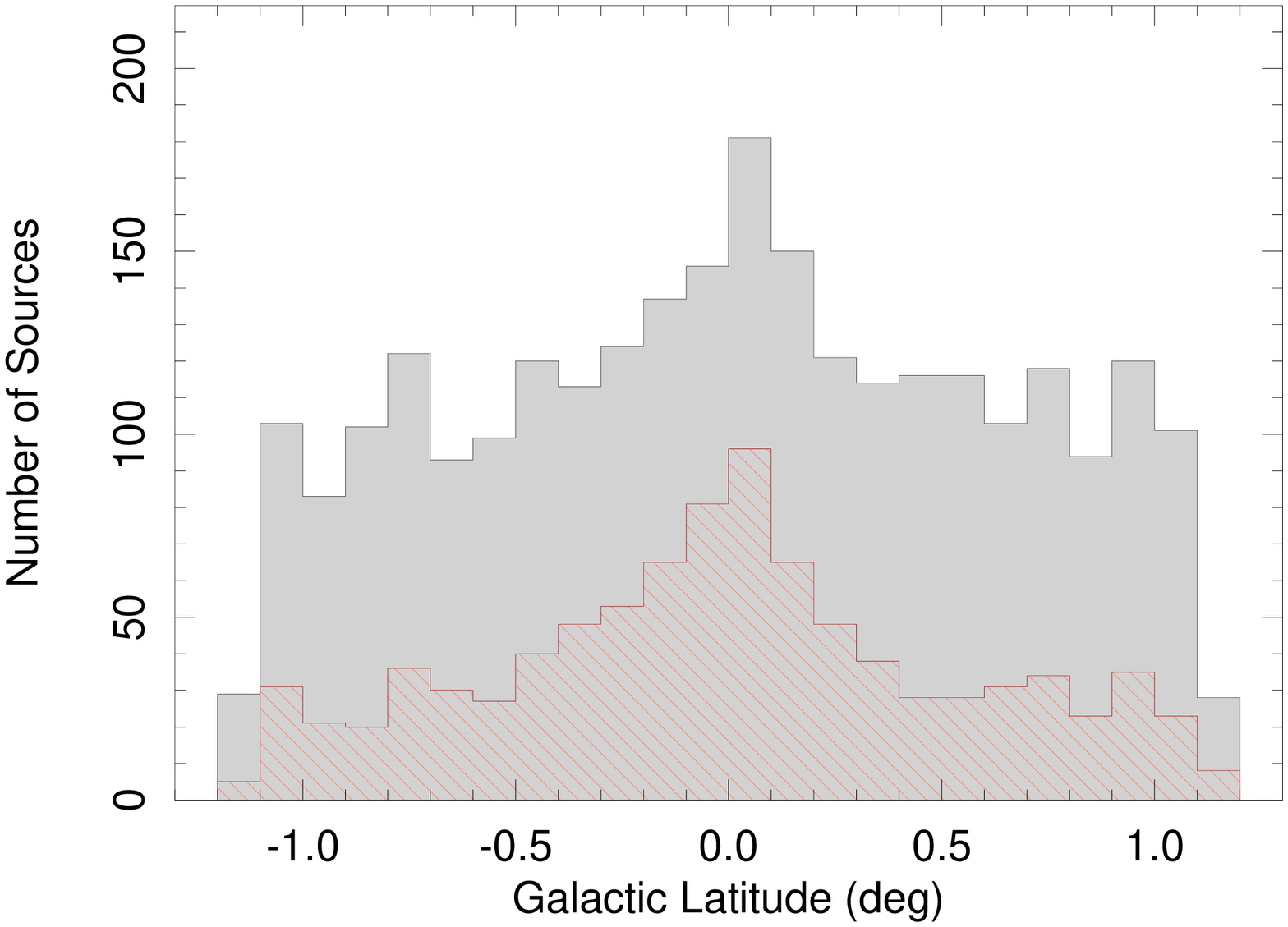}
  \includegraphics[height=7.9cm, angle=-90, trim=0 0 0 0]{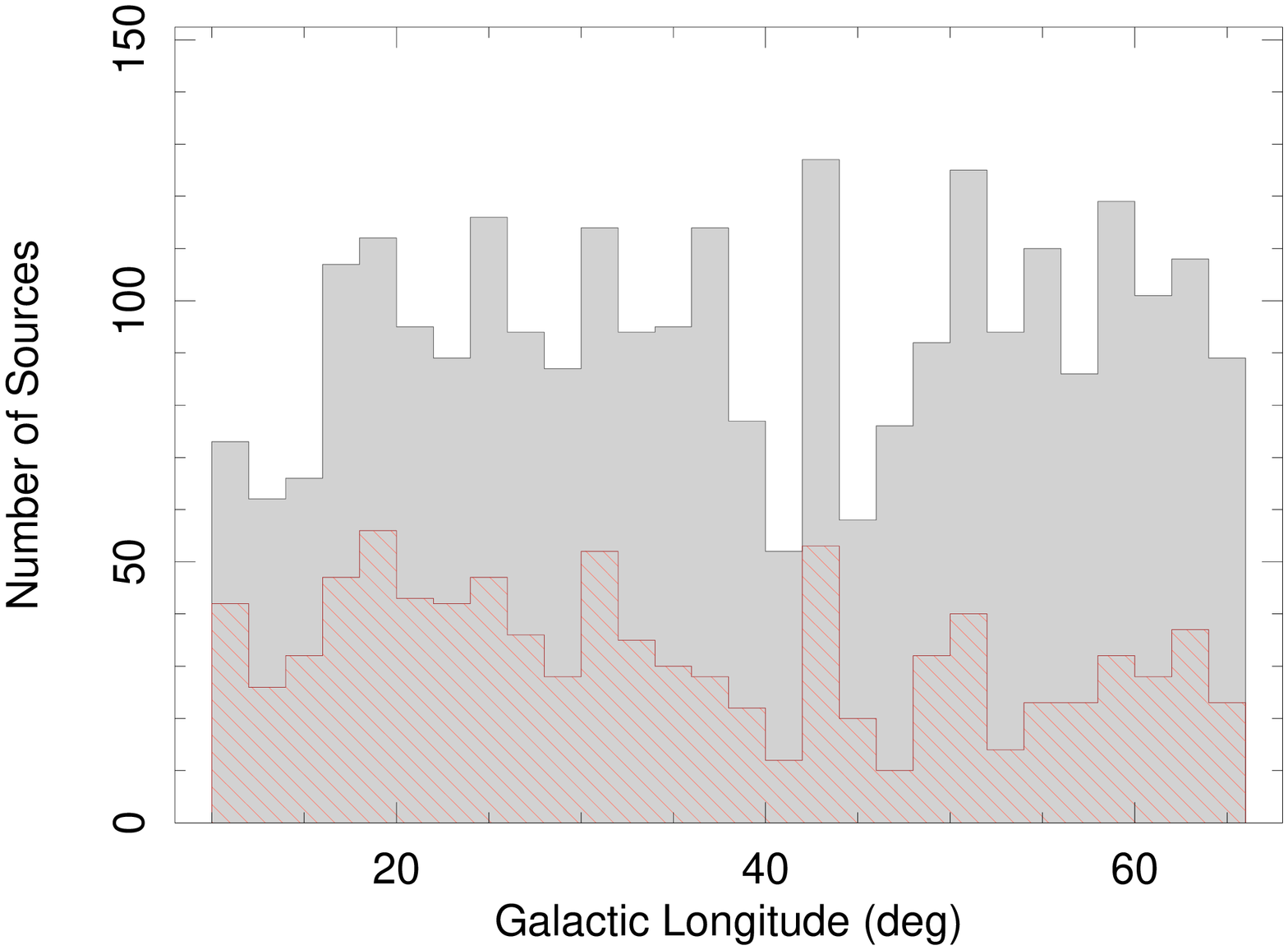}
  \end{minipage}
  \caption{\small Distribution of $\crn$ sources as a function of Galactic
    latitude ({\it upper panel}) and longitude ({\it lower panel}). In
    both panels the high-reliability catalogue ($\sigma\ge7$) is
    plotted using solid shading, while the hatched histogram contains
    a sub-sample  of resolved sources ($\theta_{\rm f}>1.8''$).}
  \label{fig:hist_galactic_distrib}
\end{figure}
\begin{figure}
  \centering
  \begin{minipage}{80mm}
  \includegraphics[height=7.9cm, angle=-90, trim=0 0 0 0]{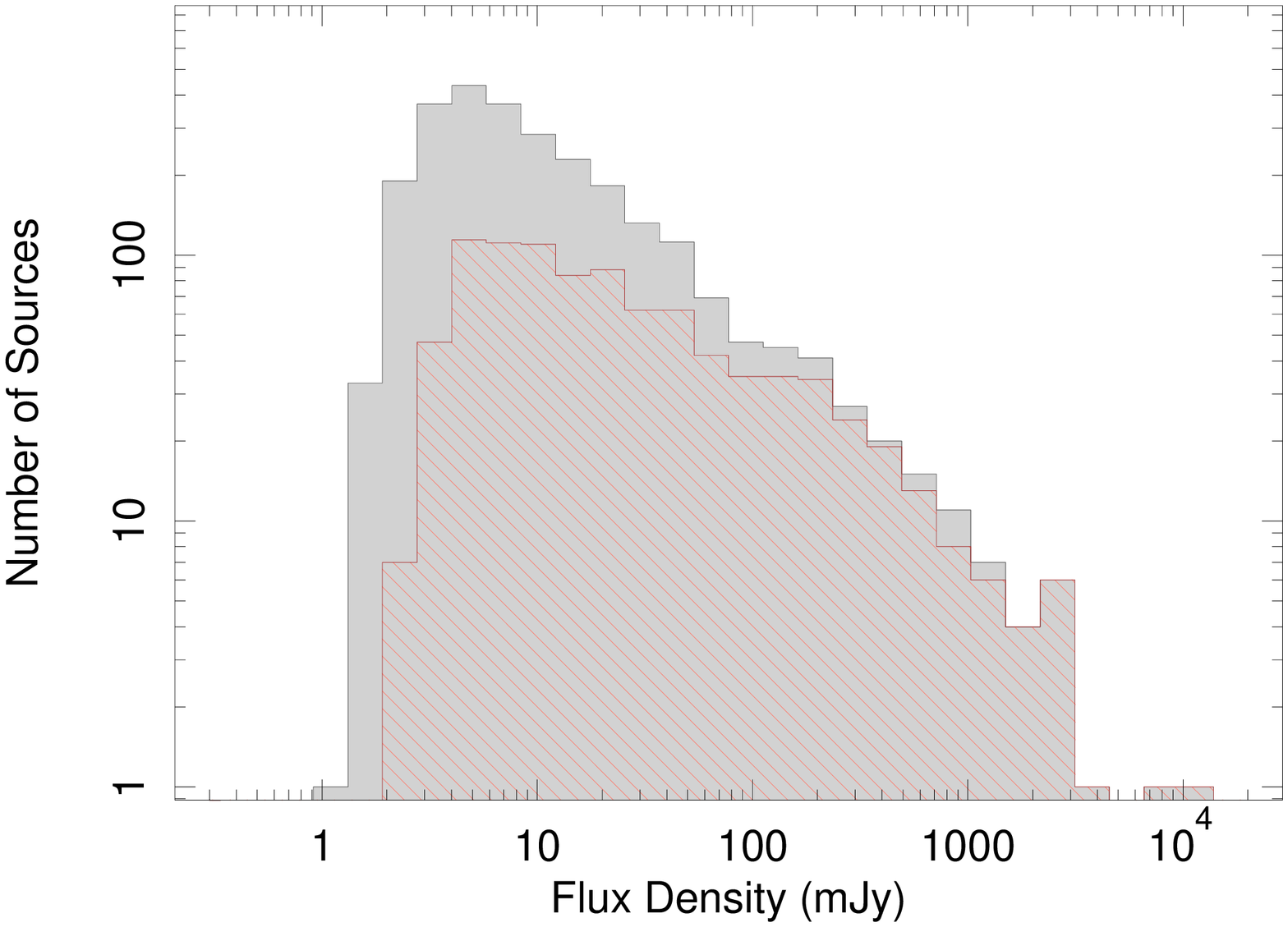}
  \includegraphics[height=7.9cm, angle=-90, trim=-20 0 0 0]{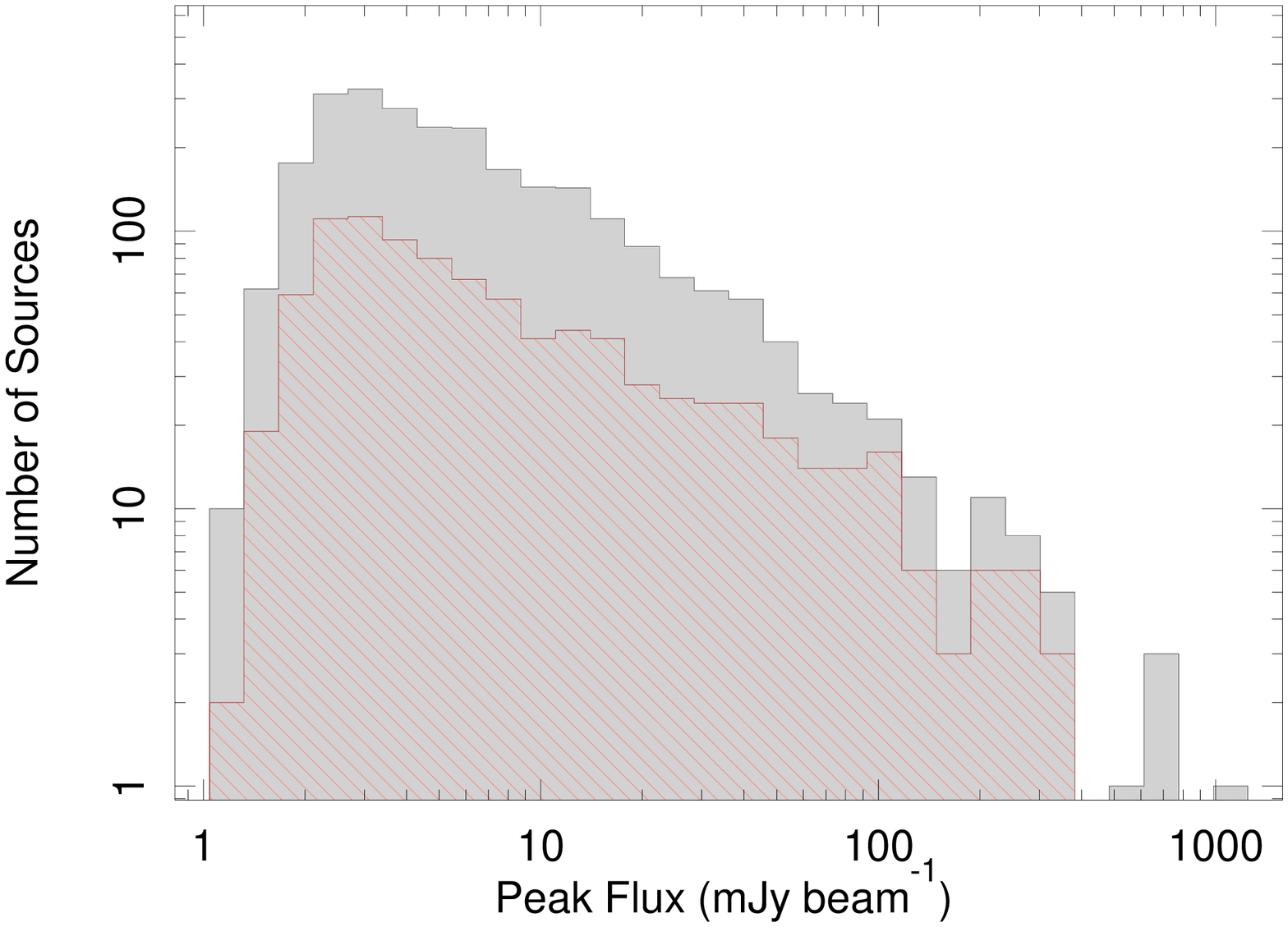}
  \end{minipage}
  \caption{\small Distribution of integrated flux density ({\it upper
      panel}) and peak flux ({\it lower panel}). The high-reliability
    catalogue is illustrated by the solid-shaded histogram, while the
    hatched histogram contains only the subset of resolved sources.}
  \label{fig:hist_flux}
\end{figure}

The full $\crn$ catalogue is offered to the astronomical community as a
plain text file or VO-table (XML based table format defined by the
Virtual Observatory) on the project
website (http://cornish.leeds.ac.uk). A query based web
interface is also available which allows the user to retrieve specific
catalogue subsets and drill down to the underlying data.


\subsection{Ensemble source properties}


\subsubsection{Angular size}
The measured angular size $\theta_{\rm f}$ quoted in the catalogue is given
by the geometric average of the major and minor fitted Gaussian FWHM axes
($\sqrt{\theta_{\rm M}\,\theta_{\rm m}}$), or by the intensity-weighted
diameter ($\theta_{\rm d}$) in the case of emission measured using a
polygonal aperture. For a bright source with a Gaussian morphology
these measurements are equivalent within the errors.
Figure~\ref{fig:hist_angscale} plots the distribution of measured
angular sizes for the high-reliability catalogue. The distribution
begins to flatten at size-scales greater than 5 arcseconds, before
tapering out at $30''$, where the upper-limit for the source-fitter is
set. Above 
sizes of $14''$ the deconvolution algorithm struggles to model the
poorly-sampled longer {\it uv}-spacings (see
Section~\ref{sec:img_extended}), 
hence the slight increase in counts at that scale - broad sources are
artificially truncated at sizes of $14''$.

The uncertainty on the angular size of $\crn$ sources is better than
$0.3''$ for 96 percent of compact ($\theta_{\rm f}<5''$) catalogue 
entries. We consider sources with $\theta_{\rm f}<1.8''$ (i.e., the
restoring beam size plus $0.3''$) to be unresolved. Sixty-one
percent of the $7\sigma$ catalogue fall into this category. Below we
examine the differences between the resolved and unresolved
populations.


\subsubsection{Galactic distribution}\label{sec:src_gal_distrib}
Figure~\ref{fig:hist_galactic_distrib} illustrates the
distribution of $\crn$ sources as a function of Galactic latitude
({\it upper panel}) and longitude ({\it lower panel}). The
solid-shaded histogram contains all sources in the high-reliability
catalogue, while the hatched histogram contains only the subset of
resolved detections (859 sources). Resolved sources account entirely
for the broad peak seen in the latitude distribution, with the
remaining unresolved detections exhibiting a flat profile. The
scale-height of the resolved latitude distribution is 
$0.47\degree$, consistent with that of $\uchii$ regions, 6.7\,GHz
methanol masers and other tracers of high-mass star-formation
(e.g.,  \citealt{Green2009},
\citealt{Urquhart2007,Urquhart2009,Urquhart2011}). The  
supposition that a large fraction of the resolved sources arise in
high-mass star-forming regions is lent weight by their
Galactic longitude distribution. The number of sources per $2\degree$
bin increases gradually towards longitude zero, while the two
spikes at $l\approx43\degree$ and $l\approx50\degree$ correspond to
the W49 and W51 complexes, respectively. Conversely, unresolved
detections (1,719 sources) show a flat distribution with Galactic
longitude and are likely to contain significant numbers of active
galactic nuclei (AGN), and other extragalactic sources. We note that
this is partly by design as course adjustments were made to the
deconvolution algorithm in order to keep the number of low-level
sources roughly constant (Figure~\ref{fig:fmrf}). The expected
density of extragalactic sources in CORNISH may be calculated from
Equation~A2 of \citet{Anglada1998}, using the 5\,GHz source counts of
\citet{Condon1984}. Assuming median values of RMS-noise for the two
regions presented in Figure \ref{fig:hist_noise}, we expect to find
$\sim2400$ extragalactic sources in our 7$\sigma$ catalogue,
consistent with the number of unresolved detections. Most
extragalactic sources can be classified as they are not detected in
any of the infrared wavebands. Specific catalogues of $\crn$ sources
identified as $\uchii$ regions, PNe and AGN will be presented in a
forthcoming paper.

\begin{figure}
  \centering
  \includegraphics[width=8.2cm, angle=0, trim=50 0 50 0]{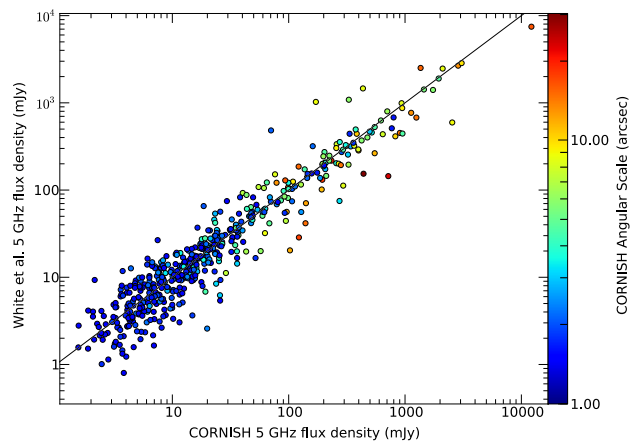}
  \caption{\small Comparison of the 5\,GHz flux density measurements
    for sources common to the $\crn$ and \citet{White2005}
    catalogues. No systematic differences are apparent in the plot,
    however, the absolute differences between the two catalogues
    increase with flux density. The points are colour-coded to show
    the angular size of the $\crn$ detections and it is clear that the
    most extended sources are responsible for the outliers seen above
    $\sim100$\,mJy.}
  \label{fig:white_flux_comp}
\end{figure}


\subsubsection{Flux density and peak flux}
Figure~\ref{fig:hist_flux} shows the distributions of flux densities
and peak fluxes for the $\crn$
sources. At flux density levels of $\sim5$\,mJy or greater the
distribution is well fitted with a power-law of index $-0.81$. Below
3\,mJy the number of sources begins to decrease as the 7$\sigma$ detection
limit is encountered. The distribution of flux densities for resolved sources  
turns over at approximately 6\,mJy due to the constraints imposed by
their selection and the $7\sigma$ signal-to-noise cutoff. 

The peak flux distributions are identical for the high-reliability and
the resolved source catalogues. Above the sensitivity cutoff
($\sim2.5$\,mJy\,beam$^{-1}$), both are fit  by a power law of 
$n_{\rm src}\propto S_{\rm 5\,GHz}^{-0.91}$.


\begin{figure*}
  \centering
  \includegraphics[width=18.0cm, angle=0, trim=0 0 0 0]{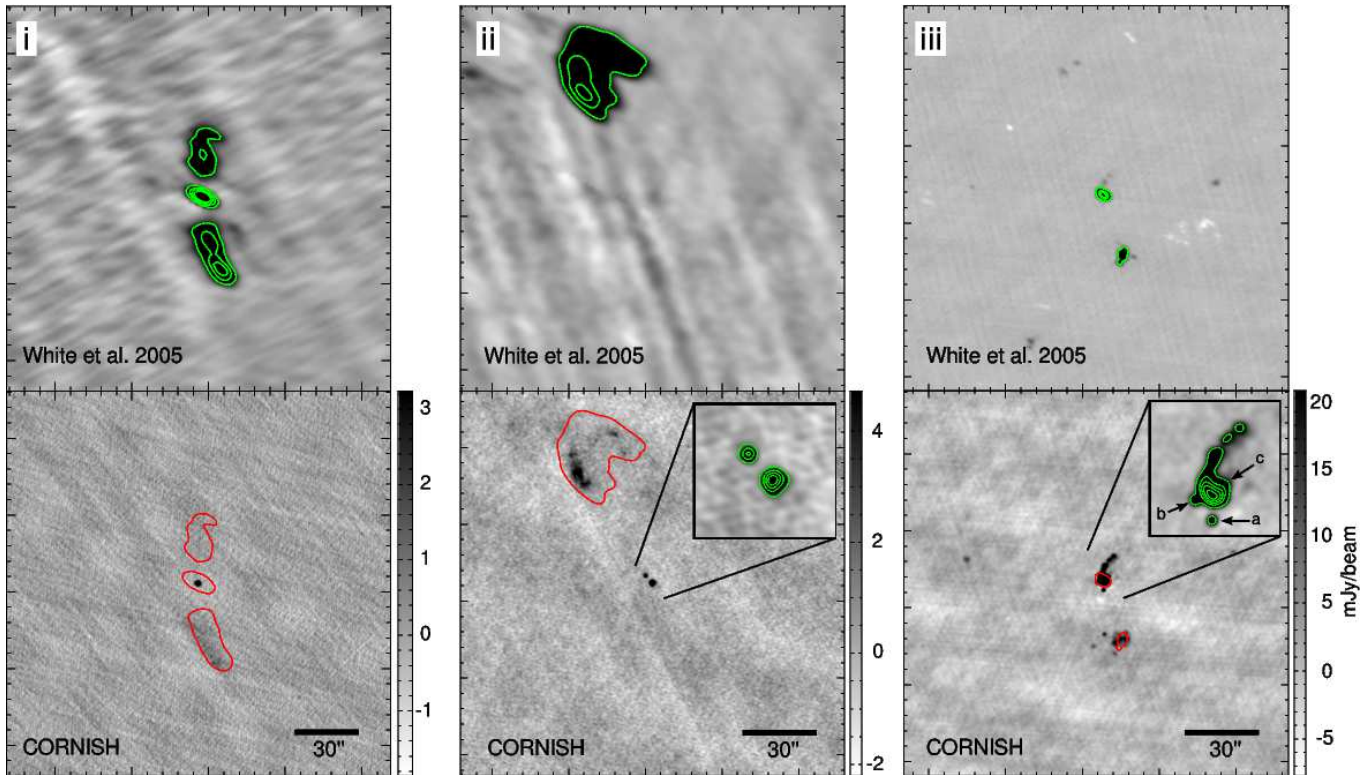}
  \caption{\small Comparison between 5\,GHz continuum images from the
    \citet{White2005} VLA survey ({\it top}) and $\crn$ ({\it
      bottom}). Panel (i) shows a radio-galaxy detected in both
    surveys. The unresolved central 
    driving source is detected in $\crn$ while the extended emission
    from the radio-lobes is resolved out. Similarly, emission on scales
    of $\sim30$ arseconds is filtered from the cometary $\hii$ region
    G24.799+0.097, shown in panel (ii). However, $\crn$ does an
    excellent job of imaging structures on scales less than $\sim14$
    arcseconds. Panel (iii) shows the star-forming region
    G34.26+0.15. The detailed structure is well-imaged by $\crn$. In
    comparison the White {\it et al.} survey struggles to resolve the
    individual knots of radio emission and significant imaging
    artefacts are present in the image. The two compact sources at the
    centre of panel (ii) are not detected in the White {\it et al.}
    survey.}
  \label{fig:white_img_comp}
\end{figure*}

\subsection{Comparison to other catalogues}
Of all prior observations the \citet{White2005} VLA survey of
the Galactic plane has the most similar observing setup and
sky-coverage. A comparison to that work serves as a  useful
sanity-check on the ensemble properties of the $\crn$ catalogue and 
images. The 5\,GHz component of the White  et al. survey was
observed using the D, DnC and C arrays. These more compact VLA
configurations (compared to B and BnA) yield better sensitivity to
extended emission than $\crn$, but at a lower resolution ($\sim6''$).
\citet{White2005} imaged the Galactic plane between 
$-10\degree<l<42\degree$ and $|b|<0.4\degree$, of which 25.6 square
degrees overlap with the $\crn$ target area. The measured noise
properties of their images are lower, with a median RMS of
$\sim0.27$\,mJy\,beam$^{-1}$ compared to $\sim0.35$\,mJy\,beam$^{-1}$ for $\crn$
data. The cutoff limit for the White  et al. source catalogue was
chosen to be $5.5\sigma$ ($\sim1.4$\,mJy\,beam$^{-1}$) 
compared to $7\sigma$ ($\sim2.5$\,mJy\,beam$^{-1}$) in this work. We would
expect similar flux densities for compact sources
($\le6''$, see Figure~\ref{fig:missing_flux}) common to both
catalogues, despite differences in {\it uv}-coverage. Systematic
errors present in either catalogue should be obvious in a flux-flux
comparison plot. 

The White et al. 5\,GHz catalogue contains 1822 entries in the
overlapping area and we match 558 of these with 521 $\crn$ sources
using a $5''$ search radius. The number of matches diminishes
significantly at matching radii greater than $2''$, however, a $5''$
matching radius was chosen to allow for offsets in the 
positions assigned to resolved sources in both catalogues.
Figure~\ref{fig:white_flux_comp} presents a comparison of the measured
flux densities for sources successfully cross-matched between the two
surveys. The measurements agree, on average, to 
within 39 percent, with no evidence of systematic differences. A
greater fraction of sources in the high flux density bins are resolved,
hence the outliers in the plot above $\sim100$\,mJy may
be attributed to imaging and measurement differences between the two
surveys. Points representing individual sources in
Figure~\ref{fig:white_flux_comp} are colour-coded to indicate angular
size in the $\crn$ catalogue. Unresolved sources (blue) cluster around
the equality line while the outliers are almost all extended (red).

In total, 1264 sources in the White et al. catalogue remain unmatched
using a simple cone search within five arcseconds. Of these, fifty
percent lie above our $7\sigma$ sensitivity threshold and are
sufficiently bright to have been detected in $\crn$.
The reasons for the disparity become apparent upon comparing the 
White et al. and $\crn$ images. A significant fraction of 
the bright, unmatched sources have angular scales greater than
$\sim20''$ in the White et al. data and are simply resolved out by the
VLA B configurations used by $\crn$. Figure~\ref{fig:white_img_comp}
({\it panels i and ii}) presents examples of such objects. The central
source powering the radio-galaxy 
shown in panel (i) is detected in both surveys as a 25\,mJy point
source. The radio-lobes have angular scales of $\sim30''$ in the White
et al. image and the brighter southern lobe has a peak flux
density of 14\,mJy\,beam$^{-1}$. In the corresponding $\crn$ image the
northern lobe is completely resolved out, while the southern lobe is
detected at a $5.5\sigma$ level and is therefore not included in the
high-reliability catalogue. Similarly, the cometary $\hii$ region
G24.799+0.097, shown in panel (ii), is resolved into multiple
components by $\crn$. When assembling the $\crn$ catalogue we took
great care to identify such over-resolved emission as a single source
(see Section~\ref{sec:measure_resolved_poly}). In the White et al.
catalogue individual Gaussian fits to complex emission are left
separate. G24.799+0.097, for example, has four catalogue entries and a
$15''$ matching radius is required to correctly match these to their
$\crn$ counterpart. 

Panel (iii) of Figure~\ref{fig:white_img_comp} presents an image of the
G34.26+0.15 star-formation region. G34.26+0.15 is divided into three
components ({\it a}, {\it b} and {\it c}), of which {\it c} is the
prototype cometary $\uchii$ region \citep{vanBuren1990}. The $\crn$
image clearly resolves all three components with excellent image
fidelity. By contrast the White et al. image barely resolves the
{\it c} component and contains significant numbers of imaging
artefacts. 

\begin{figure*}
  \centering
  \includegraphics[angle=-90, width=16.7cm, trim=30 0 0 0]{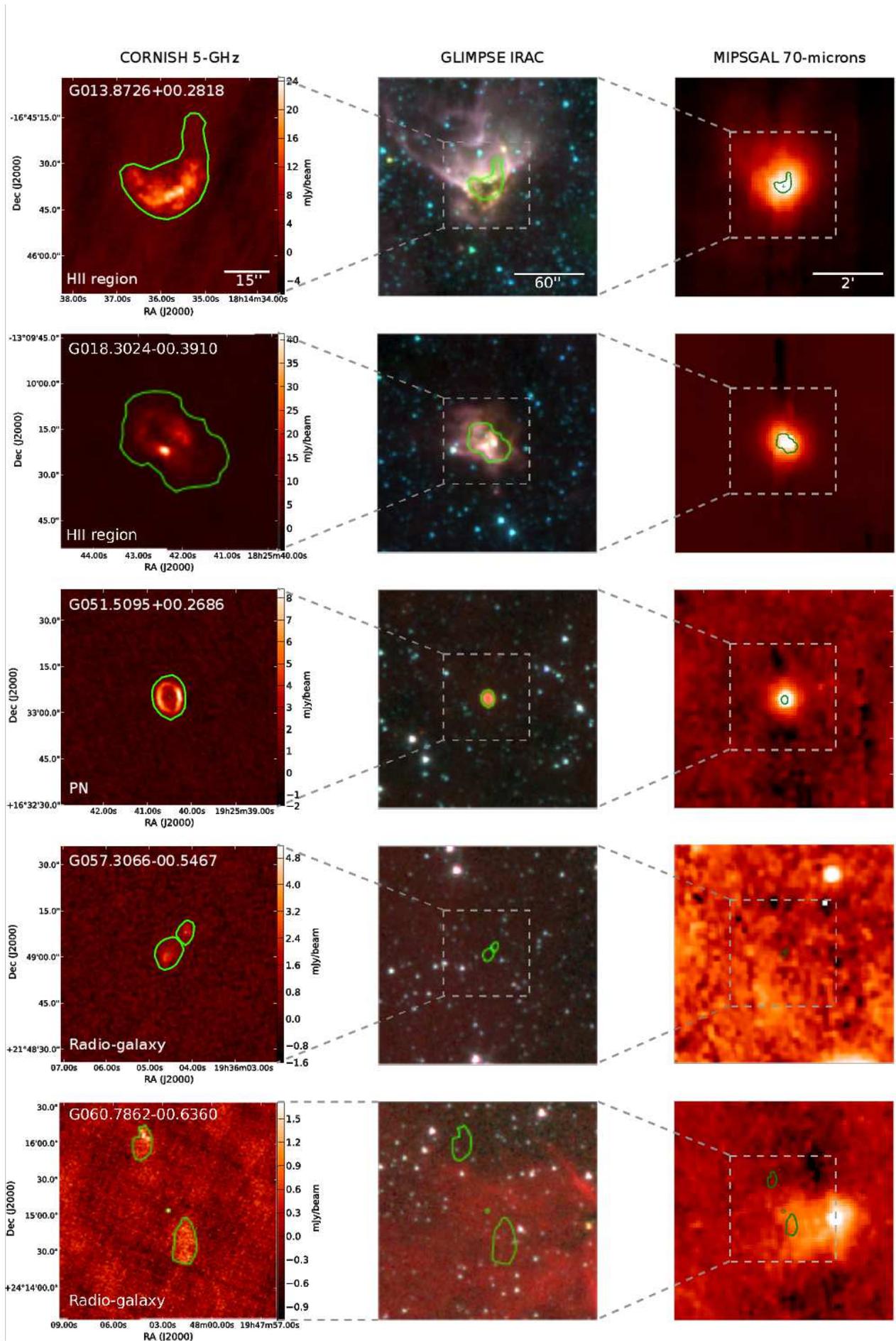}
  \caption{\small Sample $\crn$ images ({\it left column}) alongside
    {\it Spitzer} $\gli$ ({\it middle column}) and $\mip$ ({\it right
    column}) infrared data. The green polygon illustrates the aperture
    used to measure the properties of the 5\,GHz emission. In the
    three-colour $\gli$ images the 8.0\,$\micron$ band is coded red,
    the 4.5\,$\micron$ green and 3.6\,$\micron$ blue.} 
  \label{fig:example_sources}
\end{figure*}

The remaining unmatched White et al. sources derive from
intrinsic differences in the image quality between the two surveys. In
general, the $\crn$ images have more homogeneous noise properties and
are of higher quality.
The  White et al. images contain numerous compact sites of
emission not present in the equivalent $\crn$ mosaics. These often
occur in regions adjacent to very bright sources or poorly imaged
extended emission, and are themselves generally unresolved and weak
(90 percent $<10$\,mJy\,beam$^{-1}$). Their morphology and location makes them
likely to be artefacts of the imaging process. Examples of such
artefacts are visible in panel (iii) and, to a lesser extent in panel
(ii) of Figure~\ref{fig:white_img_comp}. Artefacts in the $\crn$ image
are limited to a moderate level of ripple and few noise-spikes. By
comparison the White {\it  et al.} images often contain significant
sidelobe structure and a number of spurious emission sources. The
advantages of utilising a semi-automatic pipeline are 
apparent in the high-quality of the $\crn$ images. In general, such
fields also illustrate the importance of manually inspecting the
results of automatic source-finders when dealing with under-sampled
interferometric data. 

There are 262 $\crn$ sources which have no counterpart
in the White et al. catalogue, despite peaking well above the nominal
$5.5\sigma$ detection limit. A small fraction of these are
pathological cases, like the two compact sources at the centre of the
$\crn$ image in panel (ii) of Figure~\ref{fig:white_img_comp}.  The
sources have flux densities of 12 and 20\,mJy and should be visible
in the White et al. image. A linear discontinuity cuts through the
image at this position, so we speculate that the omission derives from
a problem with the imaging process used by \citet{White2005}. The
majority of the remaining unmatched $\crn$ sources are detected below
the $10\sigma$ level and when present in the White et al. images are
washed out by ripples in the noise or other imaging artefacts.


\section{Example $\crn$ data}
The $\crn$ dataset contains objects of many types, including
$\hii$ and $\uchii$ regions, PN, evolved stars, active
binaries, radio lobes from external galaxies, and many AGN and
quasars. Figure~\ref{fig:example_sources} presents sample $\crn$
images for known objects opposite their counterpart data from the {\it
  Spitzer} $\gli$ and $\mip$ surveys. The 
first column of images shows the $\crn$ data, the second a
three-colour image made from the mid-infrared $\gli$ IRAC
bands and the third column the 70\,$\micron$ $\mip$ image. The first
two rows present examples of resolved $\hii$ regions with different
morphologies. G013.8726$+$00.2818 is a classical cometary $\hii$
region, while G018.3024$-$00.3910 is irregularly shaped. For most
resolved $\hii$ regions in $\crn$ the shape of the 5\,GHz continuum
emission is echoed and extended in the $\gli$ three-colour image. The
3.6\,$\micron$ and 8.0\,$\micron$ bands (coded blue and red,
respectively, in Figure~\ref{fig:example_sources}) contain broad lines
from poly-aromatic-hydrocarbons (PAHs), which are excited by the
strong ultraviolet radiation field \citep{Peeters2002}. Galactic
massive star-forming regions are readily identifiable via the
appearance of the filamentary PAH emission surrounding them. The
extended flocculent emission (appearing purple in the $\gli$ colour coding in
Figure~\ref{fig:example_sources}) traces the clumpy photodissociation region
(PDR) at the interface between the ionised gas and the enveloping
molecular cloud. The 
spectral-energy distribution of $\hii$ regions peaks in the
far-infrared and they are easily detected as a bright source in the
$\mip$ 70\,$\micron$ images.

The third row shows a particularly good example of a resolved
planetary nebulae (PN). The $\gli$ colours of PN are similar to those
of the $\hii$ regions, but the SED falls off more steeply in the
far-infrared, hence the $\mip$ 70\,$\micron$ band is noticeably less
bright. PNs tend to be isolated objects in the $\gli$ images, having
long since dispersed their natal molecular clouds. The expanding shell
of gas surrounding the PN also contains PAHs, which are excited by the
ultraviolet photons generated by the central stellar remnant (see
\citealt{Smith2008} and references therein). Unlike the $\hii$
regions, the PAH emission is confined to the ejected envelope
leading to simple mid-infrared morphology. G051.5095$+$00.2686 in
Figure~\ref{fig:example_sources} exhibits a similar ring-shape in both
$\crn$ and $\gli$ images. 

The final two rows present examples of radio-galaxies in which the
radio lobes have been resolved. In G057.3066$+$005467
the central driving source is not detected and the lobes are barely
resolved as two tear-drop shaped sources extended towards each 
other. The central driving source of G060.7862$-$00.6360 is
detected as a point source near the centre of the image and both
radio-lobes are well-resolved, if weak. Neither radio-galaxy has a
counterpart in any of the associated mid- or far-infrared images.


\section{Summary and future work}
The $\crn$ project has delivered the best ever complementary radio
view of the northern $\gli$ region at 5\,GHz (6-cm wavelength). With a
resolution of $\sim1.5''$ and a RMS noise level of $<0.4$\,mJy\,beam$^{-1}$,
the survey is 
tailored to search for $\uchii$ regions across the Galaxy,
but has also detected a wide range of radio-bright objects that are
also identified in other categories.

We present here a catalogue of 3,062 compact radio sources detected in
$\crn$ data above a $7\sigma$ signal-to-noise threshold. A
high-reliability subset (2,638 sources) contains has been flagged to
exclude potential spurious detections in poorly imaged {\it uv}-data. Fields
containing emission extended on scales greater than $14''$ are
poorly sampled by the {\it uv}-coverage of the VLA B-configuration, giving
rise to a small number of spurious sources. Such fields represent only two
percent of the survey area and a rigorous program of manual inspection
has flagged suspected artefacts, hence, we estimate the catalogue
reliability to be better than 99 percent. To date, the $\crn$
catalogue is the most uniformly sensitive, homogeneous and complete
list of compact radio-emission sources at 5\,GHz towards the northern
Galactic plane.

Mosaiced images and calibrated {\it uv}-data in FITS format are
available to download from the $\crn$ website
(http://cornish.leeds.ac.uk). We have created a data server,
which is operated by submitting a list of positions and serves either
postage-stamp images or calibrated {\it uv}-data. The full $\crn$
catalogue is also available online via a query based interface, as a
plain-text format file, or a VO-table. General access is also
available through the VizieR service.

Much work remains to be done in order to fully exploit the $\crn$
dataset. A future paper in the survey (Purcell et al., in prep)
will cross-match the 5\,GHz radio emission to the complementary {\it
  Spitzer} $\gli$ and UKIDSS datasets, allowing the identification of
specific source types via their SEDs.


\acknowledgments
The authors would like to thank the referee, Jim Condon, for his
thorough comments which very much improved the quality of the
paper. We would also like to thank the Director and staff of the VLA
for their assistance during the preparation of these
observations. Thanks also go to James Alison for many helpful
discussions. CRP was supported by a STFC postdoctoral grant while at
the universities of Manchester and Leeds. 

The National Radio Astronomy Observatory is a facility of the National
Science Foundation operated under cooperative agreement by Associated
Universities, Inc.

{\it Facilities:} \facility{VLA}, \facility{Spitzer (IRAC, MIPS)}


\bibliographystyle{apj}
\bibliography{cornish_cat_paper}

\begin{thebibliography}{41}
\expandafter\ifx\csname natexlab\endcsname\relax\def\natexlab#1{#1}\fi

\bibitem[{{Altenhoff} {et~al.}(1979){Altenhoff}, {Downes}, {Pauls}, \&
  {Schraml}}]{Altenhoff1979}
{Altenhoff}, W.~J., {Downes}, D., {Pauls}, T., \& {Schraml}, J. 1979, \aaps,
  35, 23

\bibitem[{{Anglada} {et~al.}(1998){Anglada}, {Villuendas}, {Estalella},
  {Beltr{\'a}n}, {Rodr{\'{\i}}guez}, {Torrelles}, \& {Curiel}}]{Anglada1998}
{Anglada}, G., {Villuendas}, E., {Estalella}, R., {et~al.} 1998, \aj, 116, 2953

\bibitem[{{Arvidsson} {et~al.}(2010){Arvidsson}, {Kerton}, {Alexander},
  {Kobulnicky}, \& {Uzpen}}]{Arvidsson2010}
{Arvidsson}, K., {Kerton}, C.~R., {Alexander}, M.~J., {Kobulnicky}, H.~A., \&
  {Uzpen}, B. 2010, \aj, 140, 462

\bibitem[{{Becker} {et~al.}(1994){Becker}, {White}, {Helfand}, \&
  {Zoonematkermani}}]{Becker1994}
{Becker}, R.~H., {White}, R.~L., {Helfand}, D.~J., \& {Zoonematkermani}, S.
  1994, \apjs, 91, 347

\bibitem[{{Briggs}(1995)}]{Briggs1995}
{Briggs}, D.~S. 1995, in Bulletin of the American Astronomical Society,
  Vol.~27, American Astronomical Society Meeting Abstracts, 112.02

\bibitem[{{Carey} {et~al.}(2009){Carey}, {Noriega-Crespo}, {Mizuno}, {Shenoy},
  {Paladini}, {Kraemer}, {Price}, {Flagey}, {Ryan}, {Ingalls}, \&
  {and~17~co-authors,}}]{Carey2009}
{Carey}, S.~J., {Noriega-Crespo}, A., {Mizuno}, D.~R., {et~al.} 2009, \pasp,
  121, 76

\bibitem[{{Churchwell} {et~al.}(2009){Churchwell}, {Babler}, {Meade},
  {Whitney}, {Benjamin}, {Indebetouw}, {Cyganowski}, {Robitaille}, {Povich},
  {Watson}, \& {Bracker}}]{Churchwell2009}
{Churchwell}, E., {Babler}, B.~L., {Meade}, M.~R., {et~al.} 2009, \pasp, 121,
  213

\bibitem[{{Condon}(1984)}]{Condon1984}
{Condon}, J.~J. 1984, \apj, 287, 461

\bibitem[{{Condon}(1997)}]{Condon1997}
---. 1997, \pasp, 109, 166

\bibitem[{{Condon} {et~al.}(1998){Condon}, {Cotton}, {Greisen}, {Yin},
  {Perley}, {Taylor}, \& {Broderick}}]{Condon1998}
{Condon}, J.~J., {Cotton}, W.~D., {Greisen}, E.~W., {et~al.} 1998, \aj, 115,
  1693

\bibitem[{{Drew} {et~al.}(2005){Drew}, {Greimel}, {Irwin}, {Aungwerojwit},
  {Barlow}, {Corradi}, {Drake}, {G{\"a}nsicke}, {Groot}, {Hales}, \&
  {Hopewell}}]{Drew2005}
{Drew}, J.~E., {Greimel}, R., {Irwin}, M.~J., {et~al.} 2005, \mnras, 362, 753

\bibitem[{{Froebrich} {et~al.}(2011){Froebrich}, {Davis}, {Ioannidis},
  {Gledhill}, {Takami}, {Chrysostomou}, {Drew}, {Eisl{\"o}ffel}, {Gosling},
  {Gredel}, \& {and~23~co-authors,}}]{Froebrich2011}
{Froebrich}, D., {Davis}, C.~J., {Ioannidis}, G., {et~al.} 2011, \mnras, 413,
  480

\bibitem[{{Giveon} {et~al.}(2005){Giveon}, {Becker}, {Helfand}, \&
  {White}}]{Giveon2005}
{Giveon}, U., {Becker}, R.~H., {Helfand}, D.~J., \& {White}, R.~L. 2005, \aj,
  129, 348

\bibitem[{{Green} {et~al.}(2009){Green}, {McClure-Griffiths}, {Caswell},
  {Ellingsen}, {Fuller}, {Quinn}, \& {Voronkov}}]{Green2009}
{Green}, J.~A., {McClure-Griffiths}, N.~M., {Caswell}, J.~L., {et~al.} 2009,
  \apjl, 696, L156

\bibitem[{{Hoare} {et~al.}(2007){Hoare}, {Kurtz}, {Lizano}, {Keto}, \&
  {Hofner}}]{Hoare2007}
{Hoare}, M.~G., {Kurtz}, S.~E., {Lizano}, S., {Keto}, E., \& {Hofner}, P. 2007,
  Protostars and Planets V, 181

\bibitem[{{Hoare} {et~al.}(2012){Hoare}, {Purcell}, {Churchwell}, \&
  {and~22~co-authors}}]{Hoare2012}
{Hoare}, M.~G., {Purcell}, C.~R., {Churchwell}, E.~B., \& {and~22~co-authors}.
  2012, submitted to PASP

\bibitem[{{Homan} \& {Lister}(2006)}]{Homan2006}
{Homan}, D.~C., \& {Lister}, M.~L. 2006, \aj, 131, 1262

\bibitem[{{Jackson} {et~al.}(2006){Jackson}, {Rathborne}, {Shah}, {Simon},
  {Bania}, {Clemens}, {Chambers}, {Johnson}, {Dormody}, {Lavoie}, \&
  {Heyer}}]{Jackson2006}
{Jackson}, J.~M., {Rathborne}, J.~M., {Shah}, R.~Y., {et~al.} 2006, \apjs, 163,
  145

\bibitem[{{Kurtz} {et~al.}(1994){Kurtz}, {Churchwell}, \& {Wood}}]{Kurtz1994}
{Kurtz}, S., {Churchwell}, E., \& {Wood}, D.~O.~S. 1994, \apjs, 91, 659

\bibitem[{{Lawrence} {et~al.}(2007){Lawrence}, {Warren}, {Almaini}, {Edge},
  {Hambly}, {Jameson}, {Lucas}, {Casali}, {Adamson}, \& {Dye}}]{Lawrence2007}
{Lawrence}, A., {Warren}, S.~J., {Almaini}, O., {et~al.} 2007, \mnras, 379,
  1599

\bibitem[{{Lucas} {et~al.}(2008){Lucas}, {Hoare}, {Longmore}, {Schr{\"o}der},
  {Davis}, {Adamson}, {Bandyopadhyay}, {de Grijs}, {Smith}, {Gosling}, \&
  {and~21~co-authors,}}]{Lucas2008}
{Lucas}, P.~W., {Hoare}, M.~G., {Longmore}, A., {et~al.} 2008, \mnras, 391, 136

\bibitem[{{Molinari} {et~al.}(2010){Molinari}, {Swinyard}, {Bally}, {Barlow},
  {Bernard}, {Martin}, {Moore}, {Noriega-Crespo}, {Plume}, {Testi}, \&
  {and~109~co-authors,}}]{Molinari2010a}
{Molinari}, S., {Swinyard}, B., {Bally}, J., {et~al.} 2010, \pasp, 122, 314

\bibitem[{{Mottram} {et~al.}(2011){Mottram}, {Hoare}, {Davies}, {Lumsden},
  {Oudmaijer}, {Urquhart}, {Moore}, {Cooper}, \& {Stead}}]{Mottram2011}
{Mottram}, J.~C., {Hoare}, M.~G., {Davies}, B., {et~al.} 2011, \apjl, 730, L33+

\bibitem[{{Peeters} {et~al.}(2002){Peeters}, {Hony}, {Van Kerckhoven},
  {Tielens}, {Allamandola}, {Hudgins}, \& {Bauschlicher}}]{Peeters2002}
{Peeters}, E., {Hony}, S., {Van Kerckhoven}, C., {et~al.} 2002, \aap, 390, 1089

\bibitem[{{Petrov} {et~al.}(2011){Petrov}, {Kovalev}, {Fomalont}, \&
  {Gordon}}]{Petrov2011}
{Petrov}, L., {Kovalev}, Y.~Y., {Fomalont}, E.~B., \& {Gordon}, D. 2011, \aj,
  142, 35

\bibitem[{{Roberts} {et~al.}(1975){Roberts}, {Cooke}, {Murray}, {Cooper},
  {Roger}, {Ribes}, \& {Biraud}}]{Roberts1975}
{Roberts}, J.~A., {Cooke}, D.~J., {Murray}, J.~D., {et~al.} 1975, Australian
  Journal of Physics, 28, 325

\bibitem[{{Robitaille} {et~al.}(2007){Robitaille}, {Cohen}, {Whitney}, {Meade},
  {Babler}, {Indebetouw}, \& {Churchwell}}]{Robitaille2007}
{Robitaille}, T.~P., {Cohen}, M., {Whitney}, B.~A., {et~al.} 2007, \aj, 134,
  2099

\bibitem[{{Schwab}(1984)}]{Schwab1984}
{Schwab}, F.~R. 1984, \aj, 89, 1076

\bibitem[{{Sewi{\l}o} {et~al.}(2011){Sewi{\l}o}, {Churchwell}, {Kurtz}, {Goss},
  \& {Hofner}}]{Sewilo2011}
{Sewi{\l}o}, M., {Churchwell}, E., {Kurtz}, S., {Goss}, W.~M., \& {Hofner}, P.
  2011, \apjs, 194, 44

\bibitem[{{Smith} \& {McLean}(2008)}]{Smith2008}
{Smith}, E.~C.~D., \& {McLean}, I.~S. 2008, \apj, 676, 408

\bibitem[{{Smith} {et~al.}(2010){Smith}, {Povich}, {Whitney}, {Churchwell},
  {Babler}, {Meade}, {Bally}, {Gehrz}, {Robitaille}, \& {Stassun}}]{Smith2010}
{Smith}, N., {Povich}, M.~S., {Whitney}, B.~A., {et~al.} 2010, \mnras, 406, 952

\bibitem[{{Steer} {et~al.}(1984){Steer}, {Dewdney}, \& {Ito}}]{Steer1984}
{Steer}, D.~G., {Dewdney}, P.~E., \& {Ito}, M.~R. 1984, \aap, 137, 159

\bibitem[{{Stil} {et~al.}(2006){Stil}, {Taylor}, {Dickey}, {Kavars}, {Martin},
  {Rothwell}, {Boothroyd}, {Lockman}, \& {McClure-Griffiths}}]{Stil2006}
{Stil}, J.~M., {Taylor}, A.~R., {Dickey}, J.~M., {et~al.} 2006, \aj, 132, 1158

\bibitem[{{Urquhart} {et~al.}(2007){Urquhart}, {Busfield}, {Hoare}, {Lumsden},
  {Clarke}, {Moore}, {Mottram}, \& {Oudmaijer}}]{Urquhart2007}
{Urquhart}, J.~S., {Busfield}, A.~L., {Hoare}, M.~G., {et~al.} 2007, \aap, 461,
  11

\bibitem[{{Urquhart} {et~al.}(2009){Urquhart}, {Hoare}, {Purcell}, {Lumsden},
  {Oudmaijer}, {Moore}, {Busfield}, {Mottram}, \& {Davies}}]{Urquhart2009}
{Urquhart}, J.~S., {Hoare}, M.~G., {Purcell}, C.~R., {et~al.} 2009, \aap, 501,
  539

\bibitem[{{Urquhart} {et~al.}(2011){Urquhart}, {Moore}, {Hoare}, {Lumsden},
  {Oudmaijer}, {Rathborne}, {Mottram}, {Davies}, \& {Stead}}]{Urquhart2011}
{Urquhart}, J.~S., {Moore}, T.~J.~T., {Hoare}, M.~G., {et~al.} 2011, \mnras,
  410, 1237

\bibitem[{{van Buren} {et~al.}(1990){van Buren}, {Mac Low}, {Wood}, \&
  {Churchwell}}]{vanBuren1990}
{van Buren}, D., {Mac Low}, M.-M., {Wood}, D.~O.~S., \& {Churchwell}, E. 1990,
  \apj, 353, 570

\bibitem[{{White} {et~al.}(2005){White}, {Becker}, \& {Helfand}}]{White2005}
{White}, R.~L., {Becker}, R.~H., \& {Helfand}, D.~J. 2005, \aj, 130, 586

\bibitem[{{White} {et~al.}(1997){White}, {Becker}, {Helfand}, \&
  {Gregg}}]{White1997}
{White}, R.~L., {Becker}, R.~H., {Helfand}, D.~J., \& {Gregg}, M.~D. 1997,
  \apj, 475, 479

\bibitem[{{Wood} \& {Churchwell}(1989)}]{Wood1989}
{Wood}, D.~O.~S., \& {Churchwell}, E. 1989, \apjs, 69, 831

\bibitem[{{Wright} {et~al.}(2010){Wright}, {Drake}, {Drew}, \&
  {Vink}}]{Wright2010}
{Wright}, N.~J., {Drake}, J.~J., {Drew}, J.~E., \& {Vink}, J.~S. 2010, \apj,
  713, 871

\end{thebibliography}

{\onecolumn
\begin{landscape}
\begin{table*}
\caption{5\,GHz sources in the CORNISH catalogue.\label{tab:cat7sigma}}
\begin{scriptsize}  
\begin{tabular}{l@{~~~}c@{~~~}c@{~~~} c@{~~~} c@{~~~} r@{~$\pm$~}r@{} c@{~~~} r@{~$\pm$~}r@{} c@{~~~} r@{~$\pm$~}r@{} c@{~~~} r@{~~} c@{} c@{~} l@{}l@{}l@{}l@{}l@{}l@{}l@{}l@{}l@{}l@{}l}
\tableline
\tableline
\rule[-0.3em]{0pt}{1.35em}(1)               & (2)              & (3)              & (4)     & (5)   & \multicolumn{2}{c}{(6)}  && \multicolumn{2}{c}{(7)} && \multicolumn{2}{c}{(8)} && (9) & (10)     & (11)   & \multicolumn{10}{c}{(12)}        \\
\rule[-0.4em]{0pt}{1.45em}Name ($l$ \& $b$) & $\alpha$ (J2000) & $\delta$ (J2000) & $\sigma_{\alpha}$ & $\sigma_{\delta}$ & \multicolumn{2}{c}{$A$} && \multicolumn{2}{c}{$S_{\rm 5\,GHz}$} &&  \multicolumn{2}{c}{$\theta_{\rm f}$} && $\theta_{\rm s}$\tablenotemark{a} & RMS     & Measure  & \multicolumn{10}{c}{Flags\tablenotemark{c}} \\
\cline{6-7} \cline{9-10} \cline{12-13}
\rule[-0.5em]{0pt}{1.55em}~~~~~~~~~~(deg) & ($^{h~m~s}$)      & ($^{\circ}~'~''$)  & ($''$)   &($''$)    & \multicolumn{2}{c}{(mJy\,bm$^{-1}$)} && \multicolumn{2}{c}{(mJy)} && \multicolumn{2}{c}{($''$)} && ($''$) &  (mJy\,bm$^{-1}$)  & type\tablenotemark{b}  &         \\
\tableline
\rule[0mm]{0pt}{1.1em}G010.1640$-$00.3655 & 18 09 27.854 & -20 19 25.99 & 0.54 (0.53) & 0.29 (0.27) & 14.25 & 2.46 (2.11) && 86.99 & 17.32 (14.89) && 3.706 & 0.454 &&3.4 & 2.03 & G &  &  & N &  &  & W &  &  & A \\
G010.4168$+$00.9356 & 18 05 09.172 & -19 28 11.15 & 0.11 (0.03) & 0.10 (0.03) & 13.00 & 1.23 (0.42) && 14.38 & 1.51 (0.78) && 1.578 & 0.043 &&~~--~~ & 0.40 & G &  &  &  &  &  &  &  &  & \\
G011.2436$+$01.0526 & 18 06 25.797 & -18 41 28.90 & 0.36 (0.35) & 0.42 (0.41) & 2.50 & 0.46 (0.40) && 11.90 & 2.62 (2.26) && 3.272 & 0.441 &&2.9 & 0.42 & G &  &  &  &  &  &  &  &  & \\
G011.3441$-$00.0381 & 18 10 40.146 & -19 07 57.14 & 0.10 (0.01) & 0.10 (0.01) & 25.34 & 2.28 (0.36) && 25.34 & 2.34 (0.64) && 1.500 & 0.027 &&~~--~~ & 0.35 & G & C &  &  &  &  &  & 7 &  & \\
G011.7210$-$00.4916 & 18 13 07.156 & -19 01 12.34 & 0.17 (0.13) & 0.17 (0.13) & 3.76 & 0.52 (0.40) && 5.93 & 1.13 (0.94) && 1.883 & 0.157 &&1.1 & 0.37 & G &  &  &  &  &  &  &  & 5 & \\
G011.7434$-$00.6502 & 18 13 45.167 & -19 04 34.68 & 0.16 (0.13) & 0.16 (0.13) & 16.97 & 1.52 (0.28) && 156.88 & 14.98 (4.16) && 4.324 & 0.008 &&4.1 & 0.28 & P &  &  &  &  &  & W &  &  & \\
G011.9368$-$00.6158 & 18 14 01.100 & -18 53 23.53 & 0.10 (0.03) & 0.10 (0.03) & 163.93 & 14.59 (0.35) && 1155.90 & 105.38 (11.79) && 5.886 & 0.003 &&5.7 & 0.35 & P &  &  &  & S &  & W &  &  & \\
G013.7481$+$00.2673 & 18 14 23.833 & -16 52 37.43 & 0.26 (0.24) & 0.24 (0.22) & 4.25 & 0.56 (0.41) && 22.93 & 3.36 (2.59) && 3.483 & 0.270 &&3.1 & 0.48 & G &  &  & N & S &  &  &  &  & A \\
G014.2365$+$00.2117 & 18 15 34.470 & -16 28 28.00 & 0.17 (0.14) & 0.17 (0.13) & 5.74 & 0.63 (0.37) && 24.34 & 2.92 (1.91) && 3.088 & 0.157 &&2.7 & 0.38 & G &  &  &  &  &  &  &  &  & \\
G014.2460$-$00.0728 & 18 16 38.295 & -16 36 06.15 & 0.50 (0.49) & 0.50 (0.49) & 11.40 & 1.04 (0.39) && 51.26 & 6.18 (3.96) && 3.935 & 0.026 &&3.6 & 0.39 & P &  &  &  &  &  &  &  &  & \\
G015.1577$-$00.0401 & 18 18 19.203 & -15 47 01.43 & 1.68 (1.67) & 1.67 (1.67) & 7.08 & 0.70 (0.42) && 12.77 & 2.32 (1.98) && 2.837 & 0.064 &&2.4 & 0.42 & P &  &  &  &  &  &  &  &  & \\
G017.7794$-$00.0082 & 18 23 18.623 & -13 27 21.64 & 0.12 (0.07) & 0.12 (0.06) & 4.66 & 0.52 (0.32) && 4.66 & 0.70 (0.56) && 1.500 & 0.081 &&~~--~~ & 0.30 & G &  &  &  &  &  &  &  &  & \\
G019.6087$-$00.2351 & 18 27 38.266 & -11 56 36.07 & 0.10 (0.02) & 0.10 (0.02) & 153.16 & 13.63 (0.56) && 2900.88 & 260.93 (26.36) && 13.108 & 0.003 &&13.0 & 0.56 & P & C &  & N & S &  &  & 7 &  & \\
G020.3969$-$00.7921 & 18 31 08.707 & -11 30 16.21 & 0.11 (0.04) & 0.11 (0.04) & 10.68 & 1.00 (0.32) && 18.18 & 1.82 (0.82) && 1.957 & 0.052 &&1.3 & 0.31 & G & C &  &  &  &  &  & 7 &  & \\
G023.0883$+$00.2242 & 18 32 32.290 & -08 38 53.38 & 0.10 (0.01) & 0.10 (0.01) & 28.75 & 2.58 (0.34) && 34.06 & 3.10 (0.67) && 1.633 & 0.027 &&~~--~~ & 0.33 & G &  &  &  & S &  &  &  &  & \\
G024.3973$+$00.7938 & 18 32 56.271 & -07 13 26.84 & 0.12 (0.07) & 0.13 (0.08) & 5.50 & 0.63 (0.40) && 6.83 & 1.03 (0.80) && 1.671 & 0.094 &&~~--~~ & 0.38 & G &  &  &  &  &  &  &  &  & \\
G025.0079$+$01.0336 & 18 33 12.925 & -06 34 18.37 & 0.16 (0.12) & 0.19 (0.16) & 2.94 & 0.40 (0.30) && 5.60 & 0.99 (0.81) && 2.071 & 0.166 &&1.4 & 0.28 & G &  &  &  &  &  &  &  &  & \\
G028.1985$-$00.0503 & 18 42 58.139 & -04 14 04.87 & 0.19 (0.17) & 0.18 (0.16) & 33.62 & 3.00 (0.36) && 136.26 & 12.94 (3.35) && 3.048 & 0.007 &&2.7 & 0.36 & P & C &  &  & S &  &  &  &  & \\
G029.0447$-$00.5989 & 18 46 28.464 & -03 43 58.01 & 1.10 (1.09) & 1.08 (1.08) & 8.07 & 0.79 (0.46) && 20.21 & 2.84 (2.16) && 2.752 & 0.039 &&2.3 & 0.46 & P &  &  & N &  &  &  &  &  & \\
G030.8000$-$01.0444 & 18 51 16.124 & -02 22 24.81 & 0.15 (0.12) & 0.23 (0.21) & 2.50 & 0.38 (0.31) && 3.66 & 0.82 (0.70) && 1.814 & 0.183 &&1.0 & 0.30 & G &  &  &  &  &  &  &  &  & \\
G032.9686$-$00.4681 & 18 53 10.271 & -00 10 50.78 & 0.10 (0.01) & 0.10 (0.01) & 87.52 & 7.80 (0.31) && 87.52 & 7.81 (0.53) && 1.500 & 0.022 &&~~--~~ & 0.32 & G &  &  &  &  &  &  &  &  & \\
G033.4543$-$00.6149 & 18 54 34.789 & 00 11 04.31 & 0.10 (0.01) & 0.10 (0.01) & 55.24 & 4.92 (0.28) && 75.82 & 6.78 (0.61) && 1.757 & 0.026 &&~~--~~ & 0.27 & G &  &  &  &  &  &  &  &  & \\
G035.3304$+$00.6133 & 18 53 37.776 & 02 24 51.19 & 0.10 (0.03) & 0.10 (0.03) & 11.69 & 1.09 (0.32) && 11.69 & 1.19 (0.56) && 1.500 & 0.037 &&~~--~~ & 0.31 & G & C &  &  &  &  &  &  &  & \\
G036.6579$-$00.0417 & 18 58 23.408 & 03 17 47.40 & 0.11 (0.05) & 0.11 (0.05) & 6.12 & 0.63 (0.32) && 6.12 & 0.78 (0.55) && 1.500 & 0.061 &&~~--~~ & 0.32 & G & C &  &  &  &  &  &  &  & \\
G040.4633$+$00.9671 & 19 01 46.445 & 07 08 31.37 & 0.11 (0.05) & 0.11 (0.05) & 7.36 & 0.75 (0.37) && 7.36 & 0.93 (0.63) && 1.500 & 0.059 &&~~--~~ & 0.35 & G &  &  &  & S &  &  &  &  & \\
G043.1684$+$00.0087 & 19 10 14.145 & 09 06 15.01 & 0.53 (0.52) & 0.52 (0.51) & 78.74 & 7.13 (1.69) && 185.25 & 21.98 (11.55) && 2.305 & 0.018 &&1.8 & 1.69 & P & C &  & N & S &  & W & 7 &  & \\
G043.1701$+$00.0078 & 19 10 14.716 & 09 06 19.18 & 0.14 (0.10) & 0.15 (0.11) & 90.46 & 8.08 (1.70) && 1108.08 & 103.11 (32.49) && 7.533 & 0.010 &&7.4 & 1.70 & P & C &  & N & S &  & W & 7 & 5 & \\
G045.2343$-$01.1294 & 19 18 13.261 & 10 24 18.43 & 0.10 (0.03) & 0.10 (0.02) & 13.37 & 1.23 (0.32) && 16.14 & 1.57 (0.63) && 1.648 & 0.037 &&~~--~~ & 0.31 & G &  &  &  &  &  &  &  &  & \\
G048.8229$+$00.5618 & 19 18 58.546 & 14 22 03.89 & 0.15 (0.11) & 0.15 (0.11) & 2.04 & 0.29 (0.23) && 2.04 & 0.46 (0.39) && 1.500 & 0.134 &&~~--~~ & 0.23 & G &  &  &  &  &  &  &  &  & \\
G048.8900$-$00.7569 & 19 23 54.299 & 13 48 23.21 & 0.14 (0.10) & 0.19 (0.17) & 4.24 & 0.55 (0.40) && 7.30 & 1.25 (1.02) && 1.969 & 0.149 &&1.3 & 0.37 & G &  &  &  &  &  &  &  &  & \\
G050.5722$-$01.1229 & 19 28 32.225 & 15 06 44.73 & 0.16 (0.13) & 0.16 (0.13) & 2.07 & 0.32 (0.26) && 2.07 & 0.53 (0.46) && 1.500 & 0.156 &&~~--~~ & 0.27 & G &  &  &  &  &  &  &  &  & A \\
G051.9859$+$00.8436 & 19 24 07.840 & 17 17 27.81 & 0.11 (0.05) & 0.11 (0.05) & 5.35 & 0.54 (0.26) && 5.35 & 0.67 (0.45) && 1.500 & 0.058 &&~~--~~ & 0.25 & G &  &  &  &  &  &  &  &  & \\
G052.0011$-$00.5305 & 19 29 12.901 & 16 39 01.94 & 0.14 (0.09) & 0.13 (0.08) & 3.06 & 0.38 (0.27) && 3.06 & 0.55 (0.48) && 1.500 & 0.106 &&~~--~~ & 0.27 & G &  &  &  & S &  &  &  &  & \\
G055.8169$-$00.8698 & 19 38 14.930 & 19 49 27.92 & 0.11 (0.05) & 0.11 (0.05) & 4.00 & 0.42 (0.23) && 4.00 & 0.54 (0.39) && 1.500 & 0.067 &&~~--~~ & 0.23 & G &  &  &  &  &  &  &  &  & \\
G058.3595$-$00.5207 & 19 42 19.210 & 22 12 30.37 & 1.61 (1.61) & 1.60 (1.59) & 2.59 & 0.30 (0.26) && 8.17 & 1.38 (1.20) && 3.106 & 0.060 &&2.7 & 0.26 & P &  &  &  &  &  &  &  &  & \\
G061.4758$+$00.0913 & 19 46 48.020 & 25 12 46.18 & 0.10 (0.01) & 0.10 (0.01) & 49.80 & 4.43 (0.46) && 3466.32 & 305.90 (29.02) && 15.978 & 0.003 &&15.9 & 0.46 & P & C &  & N & S &  &  & 7 & 5 & \\
G061.4763$+$00.0892 & 19 46 49.067 & 25 12 44.00 & 0.11 (0.04) & 0.11 (0.04) & 90.52 & 8.06 (0.46) && 718.71 & 64.37 (9.23) && 6.536 & 0.004 &&6.4 & 0.46 & P & C &  & N & S &  &  & 7 &  & \\
G062.3292$-$01.0738 & 19 53 10.033 & 25 21 21.15 & 0.10 (0.02) & 0.10 (0.02) & 11.64 & 1.07 (0.28) && 11.64 & 1.15 (0.49) && 1.500 & 0.034 &&~~--~~ & 0.30 & G & C & E &  &  &  &  &  &  & \\
G065.3071$-$00.2139 & 19 56 46.041 & 28 20 57.99 & 0.10 (0.01) & 0.10 (0.01) & 988.70 & 87.99 (0.37) && 988.70 & 88.00 (0.64) && 1.500 & 0.021 &&~~--~~ & 0.35 & G & C &  &  &  &  &  & 7 &  & \\
\tableline
\end{tabular}
\end{scriptsize}
\tablecomments{Two values are quoted for the
  uncertainty on all parameters. The first value is the absolute
  uncertainty, including measurement and calibration errors. The
  second value, in parentheses, is the uncertainty on the measurement
  alone.}
\tablenotetext{a}{Sources with $\theta_{\rm s}<1.8''$ are considered
  unresolved in the catalogue. We note that the 1.8$''$ limit is
only $\sim2$ sigma from $1.5''$ for the weakest sources, which may
cause some weak and unresolved sources to be labelled as resolved}
\tablenotetext{b}{The flux density of sources marked with an `P' in
  column (11) was measured using polygonal apertures drawn by hand on the
  images, while  a `G' means the flux density and peak-flux
  measurements were taken directly from the Gaussian fit.}
\tablenotetext{c}{The flag codes in column (12) have the following
  meanings: C\,=\,the source is part of a cluster, i.e., within 12$''$
  of another source; E\,=\,the source is within two arcminutes of a survey
  edge; N\,=\,the source lies within a high-noise region
  (RMS$>0.45$\,mJy); B\,=\,the source lies within 3$'$ of a bright
  (0.5\,Jy) source; W\,=\,{\it uv}-data for one or more fields
  contributing to a source was imaged using a smoothed weighting
  scheme; 7\,=\,the source overlaps with another 7$\sigma$ catalogue
  source; 5\,=\,the source overlaps with a 5\,--\,7$\sigma$ source;
  S\,=\,the source is located in a region with a high concentration of
  5\,--\,7$\sigma$ sources.}
\end{table*}

\end{landscape}
}

\end{document}